\documentclass[aps,reprint,superscriptaddress,nofootinbib]{revtex4-2} 
\newif\ifarxiv 
\usepackage{natbib}
\usepackage{graphicx}
\usepackage{float}
\usepackage{amsfonts,amstext}
\usepackage{url}
\usepackage{xcolor}
\usepackage{braket}
\usepackage{soul}
\usepackage{amsthm}
\usepackage{physics}
\usepackage{hhline}
\usepackage{bbm}
\usepackage{enumerate}
\usepackage{comment}
\usepackage{colortbl}
\usepackage{amssymb}
\usepackage{comment}
\usepackage{float}
\usepackage{placeins}
\usepackage[colorlinks=true,linkcolor=blue,citecolor=blue,urlcolor=blue]{hyperref}
\usepackage{cleveref}
\usepackage{ulem,nicefrac,dsfont,bm}
\usepackage{booktabs}

\newcommand{\cM}{\mathcal{M}}

\newcolumntype{C}[1]{>{\centering\arraybackslash}m{#1}}

\renewcommand{\emph}[1]{\textit{#1}}

\newtheoremstyle{theorem}
	{6pt}
	{}
	{\itshape}
	{}
	{\bfseries}
	{:}
	{.5em}
	{}
\theoremstyle{theorem}

\newtheorem{example}{Example}[section]

\newtheorem{witness}{Witness}

\newcommand{\ChJa}{\text{Choi-Jamio{\l}kowski }}

\begin{document}

\title{Quantifying the fidelity of a quantum memory}

\author{Victor Barizien}
\affiliation{
 Department of Applied Physics, University of Geneva, Switzerland}

\author{Sophie Egelhaaf}
\affiliation{
 Department of Applied Physics, University of Geneva, Switzerland}
 \affiliation{Atominstitut, Technische Universit{\"a}t Wien, Stadionallee 2, 1020 Vienna, Austria}
 \affiliation{Vienna Center for Quantum Science and Technology (VCQ), Technische Universit{\"a}t Wien, 1020 Vienna, Austria}
 
\author{Angelo Gelmini Rodriguez}
\affiliation{
 Department of Applied Physics, University of Geneva, Switzerland}

\author{Louis Nicolas}
\affiliation{
 Department of Applied Physics, University of Geneva, Switzerland}

\author{Th\'{e}o Sanchez Mejia}
\affiliation{
 Department of Applied Physics, University of Geneva, Switzerland}

\author{Mikael Afzelius}
\affiliation{
 Department of Applied Physics, University of Geneva, Switzerland}
 \affiliation{
 Geneva Quantum Centre, University of Geneva, Switzerland}

\author{Pavel Sekatski}
\author{Nicolas Brunner}
\affiliation{
 Department of Applied Physics, University of Geneva, Switzerland}
 \affiliation{
 Geneva Quantum Centre, University of Geneva, Switzerland}

\date{\today}

\begin{abstract}
Quantum memories are key ingredients for future quantum networks. Developing methods for their certification is therefore an important, yet challenging, problem. In this work, we present a general framework for certifying quantum memories and quantifying their performance in an operational and efficient manner, taking advantage of developments in the theory of quantum channels. We start by giving a pedagogical introduction to these concepts and tools and illustrate their natural application to quantum memories. Specifically, we quantify the fidelity of a quantum memory via the notion of average channel fidelity, which can be estimated precisely in an experiment via a finite set of probe states forming a so-called $2$-design. We then present a number of witnessing techniques for certifying this quantity, and show that it gives access to other important properties of the channel, such as its dimensionality and its capacity to transmit quantum information. We consider mostly a simple prepare-and-measure scenario, but discuss also the case of an entanglement-based setup, as well as a semi-device-independent approach. Finally, we apply these methods to a solid-state quantum memory experiment, certifying an average channel fidelity of 98.6\%. Our certification methods can be readily applied to other quantum devices, such as fibers, transducers and gates. 
    
\end{abstract}

\maketitle

\section{Introduction}

Quantum memories are complex physical devices, essential to future quantum communication networks \cite{Wehner2018, Lei2023}. It is thus of central importance to certify their proper functioning, and quantify their performance in an operational manner, which allows for the comparison between different implementations, based on different physical platforms.
Quantum memories are characterized by a number of features, like the efficiency, the storage-time, the multimode-capacity, or the operating wavelength \cite{Lei2023}. 
Despite the varied implementations and features, quantum memories serve a unified function: to store quantum information and mitigate decoherence. It is precisely this ability to faithfully transmit quantum information between the input and the delayed output of the memory, i.e.~to approximate the identity channel,  that we wish to certify.

In experimental works on quantum memories, various approaches are used to characterize the quantum nature of the storage process. A common procedure consists in evaluating how a set of specific input states, typically chosen ad hoc, are transmitted through the memory \cite{Specht2011single,zhang2011preparation,gundo2012quantum,vernaz2018highly,
Jiang2019,Teller2025,luo2026high}. Another approach, on the more conservative side, consists in verifying that the memory can be used to store entanglement; by sending half of an entangled state through the memory, and testing e.g.~a Bell inequality violation \cite{Ding2015, Kutluer2019, Zhu2026, Rodriguez2026} or an entanglement witness \cite{Luo2022,zhang2024realization,Zhu2026}. While these approaches offer a qualitative idea of how good the memory actually is, e.g.~by guaranteeing that it is an inherently quantum channel (as in the second approach), they are hard to compare and lack a solid operational motivation. 

In the present work, we present a general platform-independent framework for characterizing quantum memories, taking advantage of concepts and tools developed in quantum information theory. First, we propose a simple and operationally meaningful quantity for capturing the performance of a quantum memory. This quantity is {\it the average channel fidelity}~\cite{barnum1996general} -- the fidelity between the output and input states of the memory, averaged over all pure input states. The quantity retains the natural operational interpretation in terms of the average storage quality, but does not depend on an arbitrary choice of input states probing the memory. Importantly, even though the average channel fidelity involves an integration over the continuous set of all pure states, it can nevertheless be exactly estimated via a well-chosen finite set of probe states, forming a so-called $2$-design \cite{Hoggar1982}. We discuss in detail how this procedure can be implemented in practice in a simple prepare-and-measure setup.

We also show how the average channel fidelity can be tested in an entanglement-based setup. Here a maximally entangled state is prepared, half of it is then stored in the memory, and finally local measurements are performed. We discuss how the average channel fidelity relates to the so-called {\it Choi fidelity}~\cite{raginsky20011fidelity}, which has a direct operational interpretation in terms of entanglement storage~\cite{schumacher1996sending} -- it is given by the fidelity decrease after half of a maximally entangled state is stored and retrieved from the memory. This interpretation no longer depends on the specific choice of entanglement witness or Bell test. Overall, the average channel fidelity is thus an operationally motivated quantity, providing a unified approach to the problem of characterizing the fidelity of a quantum memory, based on solid theoretical concepts.

After presenting these concepts in detail in the context of a quantum memory experiment, we present methods for assessing the average channel fidelity based on efficient and experimentally-friendly tests. Going beyond $2$-designs, we develop a number of witnesses for lower bounding the average channel fidelity and investigate their performance depending on the number of required probe states. We provide particular focus on the case of qubit quantum memories. Yet, we also present general techniques for quantum memories storing high-dimensional systems. In particular, we show that a faithful witness can be constructed with a number of probe states as low as the dimension of the stored system. More generally, we show that the channel fidelity relates directly to other relevant notions, such as the dimensionality of the channel and its quantum capacity. All these methods are formulated in the simple prepare-and-measure scenario, but can be directly translated to entanglement-based experiments. In turn, we develop methods for lower bounding the average channel fidelity in a semi-device-independent scenario \cite{Brask2026}, where the measurement performed on the system at the output of the memory is treated as a black-box. 

Finally, to demonstrate the practical relevance of the methods, we apply them to a solid-state quantum memory experiment with a time-bin photonic qubit. We certify an average channel fidelity of 98.6\% and 98.5\% for the storage times of 10 $\mu$s and 125 $\mu$s and efficiencies of 16.6\% and 3.65\%, respectively. 

\begin{figure*}[t]
    \centering
    \includegraphics[width=\linewidth]{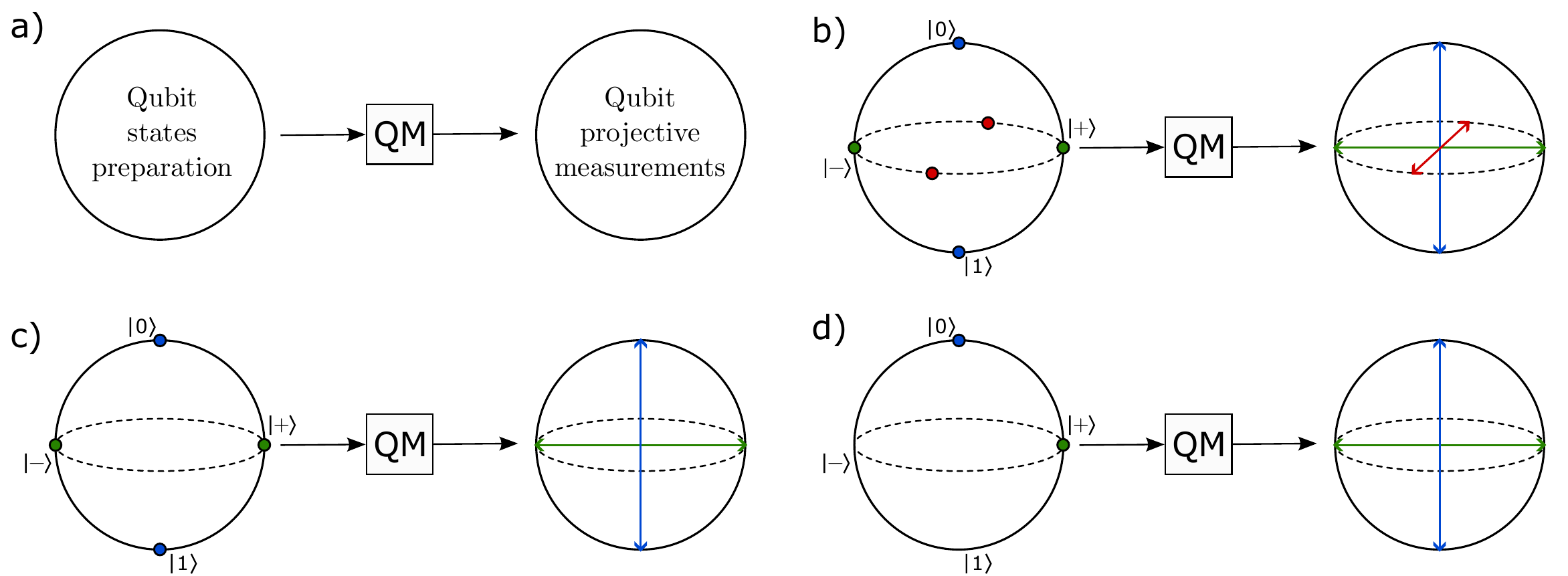}
    \caption{Estimating the average channel fidelity $F_{\rm avg}$ of a qubit memory in a prepare-and-measure setup. The general idea is presented in (a): a set of qubit states is prepared, stored in a quantum memory, and then measured via a set of projective measurements. This is illustrated with the representation of a 6 states (b), 4 states (c) and 2 states (d) strategies that can be used to characterize the fidelity of the memory via the quantity $F_{\rm avg}$. For each case, the coloured dots on the left hand side are the states (on the Bloch sphere) that need to be prepared, and the coloured arrows on the right hand side are the directions to measure on after the memory. When preparing a state of a given colour, the measurement of the same colour has to be performed. The 6-state protocol provides a direct estimate of $F_{\rm avg}$, while the other two protocols give only lower bounds on this quantity (see Section~\ref{sec:qubitwitness}). }
    \label{fig:2s}
\end{figure*}

\section{Setup and figure of merit}

A quantum memory can be formally modeled as a quantum channel $\cM$~\cite{nielsen2010quantum}, which maps an input quantum state $\rho_{\rm in}$ to an output state $\rho_{\rm out}= \cM[\rho_{\rm in}]$. An ideal quantum memory should return {\it all possible input states} unchanged, i.e. $\rho_{\rm out}= \rho_{\rm in}$, and hence corresponds to the identity channel $\cM=\rm id$. Intuitively, the performance of an actual quantum memory, described by channel $\cM$, should thus be quantified by the degree of similarity between $\cM$ and the (ideal) identity channel.

To make this intuition more concrete and quantitative, consider the task of storing a pure state $\Psi=\ketbra{\Psi}$ in the quantum memory. The similarity between the input and output state $\rho_{\text{out}}=\cM[\Psi]$ can be quantified by their fidelity $F(\rho_{\text{out}},\Psi)= \bra{\Psi} \rho_{\text{out}} \ket{\Psi}$\footnote{Here, we used the fact that the input state is pure, to write the fidelity $F(\rho,\sigma) = \left( \tr\sqrt{\sqrt{\sigma}\rho\sqrt{\sigma}}\right)^2$ as an overlap.}. In turn, a natural quantifier for the quality of the memory is to see how well it preserves input states, averaging over all possible pure states. This quantity is the {\it average channel fidelity}~\cite{barnum1996general}, which is formally given by
\begin{equation}\label{eq: avg fid}
    F_{\rm avg} (\cM, {\rm id}) := \int \dd \Psi \, F(\Psi, \cM[\Psi]),
\end{equation}
where the integration is uniform over all possible pure states (in the given Hilbert space), according to the Haar measure~\cite{Mele24}, the one measure that remains invariant under basis change.

This quantity has an intuitive and clear operational meaning, and can be directly related to physically relevant properties of the channel. Consider a memory for storing qudits (i.e.~$d$-dimensional quantum systems), then the condition $F_{\rm avg}(\cM,{\rm id})> \frac{2}{d+1}$ ensures that the memory is not entanglement-breaking. This implies that the channel is inherently quantum, as it cannot be implemented as a ``classical'' measure-and-prepare channel. Also, the condition $F_{\rm avg}(\cM,{\rm id})> \frac{1}{d}$ guarantees that the memory is not correlation-breaking, i.e.~the channel can serve to transmit some {\it classical} information, as there is non-trivial correlation between the input and output states. Moreover, the average channel fidelity also provides a lower bound for other relevant notions, such as the worst case fidelity\footnote{The worst case fidelity $F_{\rm wc} (\cM,{\rm id}) \geq (d+1)F_{\rm avg} (\cM,{\rm id}) -d$, where $F_{\rm wc}$ is defined as the quantity in Eq. \eqref{eq: avg fid} but with the integral replaced by a minimum over all input states. See \cite{castellano2025exact} for more details.}, the channel dimensionality and the quantum channel capacity (see Section~\ref{sec:dimensionality}). 

The main question is now how to estimate the average fidelity \eqref{eq: avg fid} in a real-world quantum memory experiment, where only finitely many input pure states can be tested. Perhaps surprisingly, this is not a limitation, and the quantity \eqref{eq: avg fid} can in fact be computed exactly using only certain (well-chosen) finite sets of pure probe states, forming a so-called $2$-design~\cite{Hoggar1982}. Specifically, when ${\{\Psi_x\}}_{x=1}^N$ is a 2-design one has that
\begin{equation}\label{eq: avg fid sum}
    F_{\rm avg} (\cM, {\rm id}) = {\rm T}_{\{\Psi_x\}}:=\frac{1}{N} \sum_{x=1}^N F(\Psi_x, \cM[\Psi_x]).
\end{equation}
Hence the integral in \eqref{eq: avg fid} is now replaced by a sum\footnote{Note that more generally, a $k$-design allows for the exact estimation of any polynomial up to degree $k$ on the Haar measure. Here, since the expression for the fidelity is quadratic in $\Psi_x$, a 2-design is sufficient, which has been noted in different contexts, see e.g.~\cite{bendersky2009selective,denkert2009exact}.}.

In practice, the above quantity can be efficiently estimated in the prepare-and-measure setup (see Fig.~\ref{fig:2s}a for illustration in the qubit case). Consider an optical source preparing the desired input states $\Psi_x$ (forming the $2$-design). After sending a given input state $\Psi_x$ to the memory, the output state is measured via a simple binary measurement, corresponding to a projection on the input state $\Psi_x$; i.e.~a POVM with elements $B_0 = \ket{\Psi_x}\bra{\Psi_x}$ and $B_1 = \openone - \ket{\Psi_x}\bra{\Psi_x}$. From the measurement statistics, one recovers the fidelity between the input and output states, i.e.~${\rm P} (b=0| \Psi_x) = F(\Psi_x, \cM[\Psi_x])$, where ${\rm P} (b| \Psi_x)$ denotes the probability to get outcome $b$ for input state $\Psi_x$. In turn, we get that the average fidelity is directly obtained from the measurement data via
\begin{align} \label{eq: T def}
    F_{\rm avg} (\cM, {\rm id}) =\frac{1}{N} \sum_{x=1}^N {\rm P}(0| \Psi_x).
\end{align}
Note that in general, there exists many different set of states $\{\Psi_x \}$ that form a 2-design, with varying cardinalities (i.e. number of states in the set) and symmetries; see below for some examples. Note also that there exist weighted 2-designs~\cite{roy2007weighted}, where the states $\Psi_x$ are not sampled with uniform probability.

\subsection{Qubit quantum memory} \label{sec:perfectqubit}

As an illustrative example, let us discuss in detail the case of a quantum memory for qubits, encoded e.g.~in the polarisation of single photons. 

First, we need to identify a set of qubit states forming a $2$-design. There exist several possibilities, among which the simplest is arguably the set of 6 Pauli eigenstates, i.e.
\begin{equation}
    \scalebox{0.975}{$\mathds{S}_6=\left\{\ket{0},\ket{1},\frac{\ket{0}+\ket{1}}{\sqrt{2}},\frac{\ket{0}-\ket{1}}{\sqrt{2}}, \frac{\ket{0}+i\ket{1}}{\sqrt{2}}, \frac{\ket{0}-i\ket{1}}{\sqrt{2}}\right\}$} 
\end{equation} 
the eigenstates of the Pauli matrices $\sigma_z$, $\sigma_x$ and $\sigma_y$. For each state, the output state is measured in the corresponding Pauli basis. Since the states form pairs of orthogonal states, we only need 3 projective measurements, which correspond to a complete set of mutually unbiased bases (MUBs) for the case of qubits\footnote{Note that Ref.~\cite{BOWDREY2002258} showed how the average fidelity of a qubit channel can be computed via the set $\mathds{S}_6$, i.e. recovering equation~\eqref{eq: avg fid sum}. The authors used a direct integration over the Bloch sphere, without noticing the connection to 2-designs.}. The protocol is illustrated in Fig.~\ref{fig:2s}b and consists of a simple prepare-and-measure setup. For each Pauli probe state $\Psi_x$ in $\mathds{S}_6$, one can then estimate the corresponding probability $P(0|\Psi_x)$ giving the input-output fidelity. In turn, the average channel fidelity is obtained via Eq. \eqref{eq: T def}. Of course, in a realistic quantum memory experiment, the storing process has a limited efficiency and the experiment is prone to other unavoidable sources of loss. The probabilities $P(0|\Psi_x)$ in Eq.~\eqref{eq: avg fid sum} are thus typically computed based on the post-selected (conclusive) data.

More generally, the ensemble $\mathds{S}_6$ represents a very natural and intuitive choice of probe states for testing a qubit memory. In fact, it has been used (as an ad hoc choice) in several qubit memory experiments~\cite{Specht2011single,zhang2011preparation,gundo2012quantum,vernaz2018highly,
Jiang2019,Teller2025,luo2026high}, where the input-output fidelity of the six states has been used to test the memory, however, without noticing the connection to the average channel fidelity, as given by Eq.~\eqref{eq: avg fid sum}.

Interestingly, there exist other choices for qubit 2-designs. In particular, the set $\mathds{S}_{\rm SIC}$ of 4 states forming a regular tetrahedron on the Bloch sphere and defining the so-called symmetric informationally complete (SIC) POVM~\cite{Renes_2004} is also a $2$-design. In principle, an advantage of such a set is that it requires less states\footnote{Note that the SIC POVM provides in fact the 2-design of minimal cardinality~\cite{johnson2001basic}, namely with $d^2$ states.}, but it will require four projective measurements to estimate the quantity \eqref{eq: avg fid sum}, since the states are no longer pairwise orthogonal.

\subsection{Estimating the average fidelity via entanglement}

Another approach for testing a quantum memory takes advantage of entanglement. Specifically, one first prepares an entangled state between two optical modes, sends one mode into the quantum memory, and finally measures the output state. Intuitively, if the final state is still entangled, this implies that the quantum memory could preserve the entanglement and is thus a true quantum channel. This approach has been followed in several experimental works, implementing e.g.~Bell inequality tests for verifying the output entanglement. 

This approach can also be rooted on firm grounds and serve as a precise and rigorous protocol for quantifying the quality of the quantum memory via the average channel fidelity $F_{\rm avg}$. This relies on the so-called Choi-Jamiliokowsky isomorphism~\cite{Jamiolkowski72}: to any quantum channel $\cM$ one can associate a bipartite quantum state $\varrho_\cM$ and vice-versa. For this, consider initially preparing a maximally entangled state $\Phi^+$, and sending one subsystem through the channel $\cM$. The resulting output defines the so-called Choi state associated to $\cM$, it is given by 
\begin{equation}
\varrho^\cM_{AB}:=({\rm id}_A\otimes \cM_B)[\Phi^+],
\end{equation}
where $B$ is the system that is stored in the memory and $A$ is the auxiliary system.

\begin{figure}
    \centering
    \includegraphics[width=0.9\linewidth]{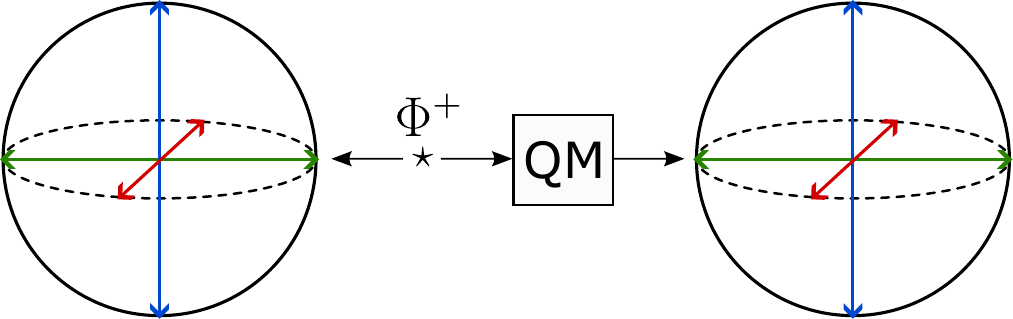}
    \caption{The average channel fidelity can also be estimated in an entanglement-based setup. A source prepares a maximally entanglement state $\Phi^+ = \ketbra{\Phi^+}$ with $\ket{\Phi^+}=(\ket{00}+ \ket{11})/\sqrt{2}$. One particle (mode) is sent to the memory. Finally, both particles (modes) are measured locally, in the same Pauli basis. From the 3 correlators $E_{SS}$ for $S\in \{X,Y,Z\}$ (respectively above in blue, red or green) one can directly compute the average channel fidelity via Eq.~\eqref{eq:Corr_fid}.}
    \label{fig:6state_ent}
\end{figure}

In turn, it is natural to quantify the degree of entanglement of this output state via its fidelity with the (ideally stored) maximally entangled state~\cite{schumacher1996sending}, as given by the quantity 
\begin{equation}
F_{\rm CJ} (\cM,{\rm id}):= \bra{\Phi^+} \varrho^\cM_{AB} \ket{\Phi^+}.
\end{equation}
Intuitively this quantity also captures the quality of the memory channel. Indeed, it is precisely the {\it Choi fidelity}~\cite{raginsky20011fidelity} between $\cM$ and the identity channel representing the ideal memory. Furthermore, since in our case one of the channels (the identity) is unitary,
the average and Choi fidelity are one-to-one related~\cite{Horodecki1999,Nielsen2002}, via
\begin{equation}\label{eq: avg vs CJ}
   (d+1) F_{\rm avg} (\cM,{\rm id})= d \, F_{\rm CJ} (\cM,{\rm id})+1.
\end{equation}
For completeness, in Appendix~\ref{app: F to F} we give a pedagogical derivation of this identity. In summary, $F_{\rm avg} (\cM,{\rm id})$ and $ F_{\rm CJ} (\cM,{\rm id})$ are mere rescalings of one another and capture the same exact property of the quantum memory. Note that experimental groups often use quantum process tomography to estimate the process fidelity of the channel, which coincides with the Choi fidelity $F_{\rm CJ} (\cM,{\rm id})$. As we discussed, the same quantity can be derived from the average channel fidelity $F_{\rm avg} (\cM,{\rm id})$, hence using fewer measurements (see Discussion).

In a practical quantum memory experiment based on entanglement, the average channel fidelity can also be directly obtained via the concept of 2-designs. Specifically, in the qubit case, each party can measure locally  in the 3 Pauli bases $\sigma_S$ with $S \in \{X,Y,Z \}$, and calculate the 3 correlators 
\begin{align}
\label{def_E_SS_corr}
   {\rm E}_{SS} := \tr \Big(\sigma_S\otimes \sigma_S\,  ( {\rm id}_A\otimes \cM_B)[\Phi^+]\Big)
\end{align}
as illustrated in Fig.~\ref{fig:6state_ent}. In turn, the average channel fidelity is given by
\begin{align}\label{eq:Corr_fid}
  F_{\rm avg}(\mathcal{M},{\rm id})  =\frac{3+{\rm E}_{XX}-{\rm E}_{YY} + {\rm E}_{ZZ}}{6} ,
\end{align}
or equivalently 
\begin{align}
  \label{F_CJ_corrs}
  F_{\rm CJ}(\mathcal{M},{\rm id})  =\frac{1+ {\rm E}_{XX}-{\rm E}_{YY} + {\rm E}_{ZZ}}{4} .
\end{align}

This last relation has been used in memory experiments to characterize the preservation of entanglement during storage \cite{Luo2022,zhang2024realization,Zhu2026}.

\subsection{Probabilistic quantum memories}

In practice, quantum memories have limited efficiency due to unavoidable technical imperfections. A typical example is a quantum memory for a photonic qubit (polarization or time-bin) which sometimes outputs vacuum. In general, such a memory can be represented by a channel of the form 
\begin{align}
    \cM[\,\cdot\,] =  {\cM_\checkmark}[\,\cdot\,] + \left(1-\tr \cM_\checkmark[\,\cdot\,]\right) \ketbra{\varnothing}.
\end{align}
The first term ${\cM_\checkmark}$ is a completely positive trace-non-increasing map describing successful storage in the memory. The second term represents the loss process, resulting in an output state $\ketbra{\varnothing}$, which is orthogonal to the input Hilbert space.

In this case, the memory can be described via two independent quantities, the average efficiency
\begin{equation}\label{eq: efficacite}
    \eta(\cM) = \int \dd \Psi \tr \left(\cM_\checkmark[\Psi] \right)=\tr \Big[({\rm id}\otimes  \cM_\checkmark)[\Phi^+] \Big]
\end{equation}
and the conditional (post-selected) fidelity
\begin{align}
    \bar F_{\rm avg}(\cM,{\rm id}) &= \frac{F_{\rm avg}(\cM,{\rm id})}{\eta(\cM)}.
\end{align}
Both quantities can be estimated via a set of probe states forming a $2$-design. For the conditional fidelity, the numerator is computed similarly to the unit efficiency case, as the contribution of the outcome $\ketbra{\varnothing}$ to the fidelity vanishes we have $F_{\rm avg}(\cM,{\rm id}) = F_{\rm avg}(\cM_\checkmark,{\rm id})= \int \dd \Psi \bra{\Psi}\cM_\checkmark[\Psi] \ket{\Psi}$. Note that one can also define the conditional Choi fidelity $ \bar F_{\rm CJ}(\cM,{\rm id}) = \frac{F_{\rm CJ}(\cM,{\rm id})}{\eta(\cM)}$, which is still related to $\bar F_{\rm avg}(\cM,{\rm id}) $ by the relation \eqref{eq: avg vs CJ} as shown in Appendix \ref{app: F to F}.

\section{Witnesses for average fidelity}

It is relevant to ask whether the average fidelity $F_{\rm avg}$ can be estimated from more general sets of states and measurements, not necessarily forming a 2-design. In this case, one expects to obtain (non-tight) lower bounds. This question is of course well motivated from the experimental point of view, where it may often be cumbersome (or even impossible) to prepare (and measure) sets forming 2-designs (as e.g.~a complete set of mutually unbiased basis). In this case, we will see that it is nevertheless possible to derive lower bounds on $F_{\rm avg}$ based on the observed statistics.

\subsection{Qubit witnesses} \label{sec:qubitwitness}

First, let us reconsider the above question (Section~\ref{sec:perfectqubit}) of certifying a qubit quantum memory. In the experiment, a set of states $\{\Psi_x\}$ are prepared, sent through the memory and finally measured with a binary measurement projecting along the respective input states. From the measured data, one can still estimate the quantity
\begin{equation}
{\rm T}_{\{\Psi_x\}} = \frac{1}{|\{\Psi_x\}|} \sum_{x=1}^{|\{\Psi_x\}|} F(\Psi_x,\cM[\Psi_x]) \,,
\end{equation}
from which we now aim to obtain a lower bound on $F_{\rm avg}$. We focus on two natural examples, illustrated in Fig.~\ref{fig:2s}c-d: (i) the set of four Pauli states on the equator of the Bloch sphere $\mathds{S}_4=\big\{\ket{0},\ket{1},\ket{+},\ket{-} \big\}$ and (ii) two non-orthogonal Pauli states $\mathds{S}_2=\big\{\ket{0},\ket{+}\big\}$. Note that in both cases, we require only two projective measurements ($X$ and $Z$). For these schemes, we can derive the following bounds 
\begin{align} \label{eq:witnessbounds}
     F_{\rm avg}(\cM,{\rm id})&\geq \begin{cases}
         (4\, {\rm T}_{\mathds{S}_4}-1)/3,\\
         f_2({\rm T}_{\mathds{S}_2}),
     \end{cases}
\end{align}
where $f_2$ is a convex function defined as the solution of a simple semi-definite program; for details, see Appendix~\ref{app:SDPopt}. In Fig.~\ref{fig:FvsW}, we plot the certified average fidelity as a function of the observed witness value. As expected, we see that both protocols give only lower bounds, and thus underestimate the average fidelity (for instance verifying that the memory is not entanglement-breaking requires ${\rm T}_{\mathds{S}_4}>3/4$ instead of ${\rm T}_{\mathds{S}_6}>2/3$). Yet they provide valid and useful bounds and require simpler experimental tests. In particular, we observe that both tests can certify an ideal memory, i.e.~observing ${\rm T}_{\{\Psi_x\}}=1$ implies $F_{\rm avg}(\cM,{\rm id})=1$. The protocol with two states ($\mathds{S}_2$) is therefore minimal, as a protocol with a single state cannot achieve this, nor show that the channel is not entanglement-breaking. 


\begin{figure}[t]
    \centering
    \includegraphics[width=0.95\linewidth]{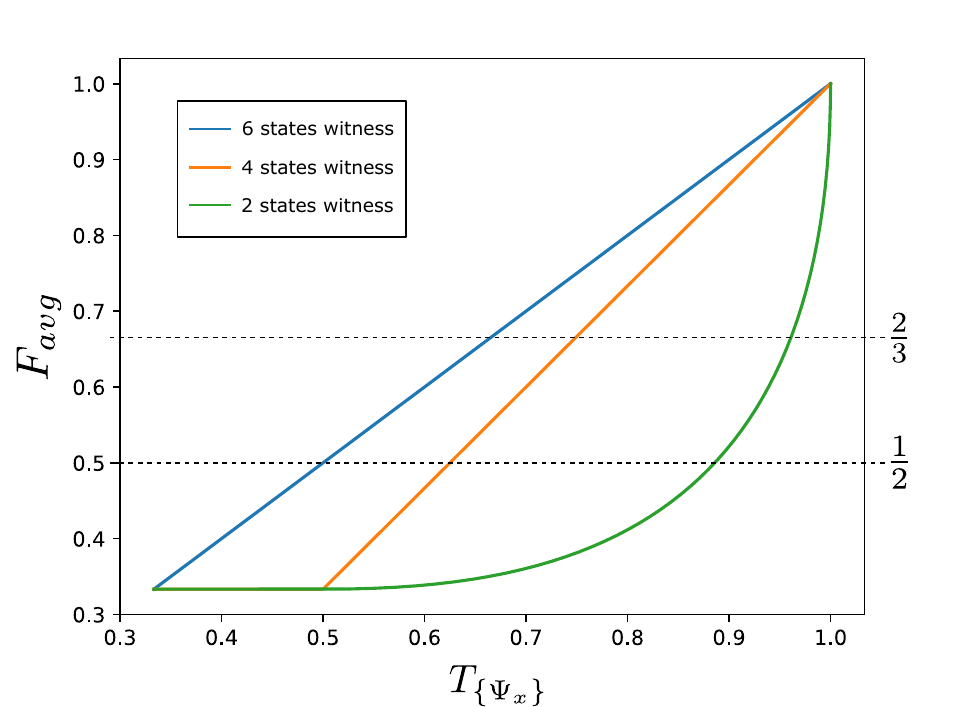}
    \caption{Lower bound on the average channel fidelity $F_{\rm avg}(\cM, {\rm id})$ of a qubit memory $\cM$ as a function of the witness value ${\rm T}_{\{\Psi_x\}}$ using 2, 4 and 6 states (respectively in green, orange and blue). For the 6-state witness, the quantity ${\rm T}_{\{\Psi_x\}}$ corresponds to $F_{\rm avg}$. The 4- and 2-state protocols provide lower bounds on $F_{\rm avg}$, which are computed via an SDP optimiasation, as described in Appendix~\ref{app:SDPopt}. Certifying a fidelity $F_{\rm avg} > 2/3$ ensures that the channel is not entanglement-breaking, while $F_{\rm avg} > 1/2$ guarantees the channel is not correlation-breaking.}
    \label{fig:FvsW}
\end{figure}

It is worth mentioning that for the witnesses, it is only the relation between the states in the sets $\mathds{S}_k$ that plays a role. Thus any six, four or two states picked from three (or two) complementary bases lead to the same bounds.

Note that in the entanglement scenario, we can also rewrite the four-state witness of Eq.~\eqref{eq:witnessbounds} in terms of the ${\rm E}_{SS}$ correlators, cf.~Eq.~\eqref{def_E_SS_corr}. Choosing the $X$ and $Z$ bases, for instance, we find
 \begin{align}
   F_{\rm avg}({\rm id},\mathcal{M}) \geq  \frac{{\rm E}_{XX} + {\rm E}_{ZZ}+1}{3}.
 \end{align}

\subsection{Witnesses for higher dimensions} \label{sec:highdim}

So far, we have focused on the case of a qubit quantum memory, and will now discuss the generalization to arbitrary finite dimensions. Crucially, note that Eq.~\eqref{eq: avg fid sum} is valid regardless of the dimension of the quantum system stored by the memory: that is, for a 2-design $\mathds{S}=\{\Psi_x\}$ the identity ${\rm T}_{\{\Psi_x\}}= F_{\rm avg} (\cM,{\rm id})$ holds (by definition of a 2-design).  

Explicit constructions of 2-designs are known for specific dimensions. Complete sets of mutually unbiased bases (MUBs)---which can be seen as a generalisation of the 3 Pauli bases for qubits---form 2-designs \cite{klappenecker2005}. These are known to exist for Hilbert space dimensions $d$ that are power prime \cite{Wootters1989}, but are conjectured not to exist in all dimensions \cite{Horodecki2022}. They consist of $d(d+1)$ states. Another construction for 2-designs is via the generalisation of the SIC POVM. The latter consist of $d^2$ states---the minimal cardinality for a 2-design---and are conjectured to exist for any dimension \cite{Horodecki2022}, but are only constructed explicitly for dimensions up to about $d=67$ \cite{Scott2010}. Note also, that weighted 2-designs can always be constructed from {\it sufficiently many} orthonormal bases~\cite{roy2007weighted}.

Nevertheless, just like in the case of qubits, it is natural to develop witnesses for lower bounding the average fidelity that are based on sets of states that do not form a 2-design, in particular considering sets with less than $d^2$ states. Below we present three constructions of such witnesses. For these examples, it is easier to obtain bounds on the \ChJa fidelity $F_{\rm CJ}(\mathcal{M},{\rm id})$, which is directly related to the average fidelity via Eq.~\eqref{eq: avg vs CJ}.\\

As our first example, we consider a memory storing $n$ qubits simultaneously, e.g.~storing two (or more) degrees of freedom carried by a photon, like polarization and time-bin \cite{Tiranov2015,Zhang2016}, or an atomic-ensemble memory that simultaneously stores several photonic qubits in different spatial modes \cite{Jiang2019, zhang2024realization,Teller2025}. Notice that if the memory channel is ideal for each of the $n$ qubit subsystems, it must also be ideal for the global system, i.e.~it is a $2^n$-dimensional identity channel. Thus, intuitively, combining fidelity witnesses for each subsystem should provide a witness for the global fidelity of the memory. Indeed, starting from 6 and 4-qubit-state sets, it can be shown that the global fidelity satisfies
\begin{align}
    F_{\rm CJ}(\mathcal{M},{\rm id}) &= \frac{3}{2}T_{\mathds{S}_6^{\otimes n}}-\frac{1}{2} \geq 2 \, T_{\mathds{S}_4^{\otimes n}}-1,
\end{align}
where 
\begin{align}
   T_{\mathds{S}_k^{\otimes n}} = \frac{1}{| \mathds{S}_k|^n}\!\!\sum_{\Psi_{x_1},..,\Psi_{x_n}\in \mathds{S}_k} \!\! F\left(\bigotimes_{i=1}^n \Psi_{x_i}, \!\mathcal{M}\Big[\bigotimes_{i=1}^n \Psi_{x_i}\Big]\right).
\end{align}
In words, in order to lower bound the fidelity  of the $2^n$-dimensional memory it is sufficient to estimate the average fidelity over all product states obtained by independently choosing a state from $\mathds{S}_4$ or $\mathds{S}_6$ for each qubit. Notice that here the total number of required probe states is $6^n$ or $4^n$, which is of the order $O(2^{2n})$, as for a 2-design. Nevertheless the practical advantage can come from the fact that the $n$ qubits only need to be controlled independently. The following result provides such bounds for general composite systems and linear witnesses.

\begin{witness}[Product witness] \label{wit:product}Consider a memory channel $\cM$ for a system of product dimension $d=\prod_{i=1}^n d_i$. For each subsystem of dimension $d_i$ consider a set of states $\mathds{S}_i$ and a linear fidelity witness
\begin{align}
F_{\rm CJ}(\mathcal{M}_i,{\rm id}) &\geq (1+\delta_i)T_{\mathds{S}_i} - \delta_i \end{align}
with $T_{\mathds{S}_i} :=\frac{1}{|\mathds{S}_i|}\sum_{\Psi_x \in \mathds{S}_i} \tr (\Psi_x \mathcal{M}_i[\Psi_x])$. 
Then for the product set $\mathds{S}:=\left\{\bigotimes_{i=1}^n \Psi_{x_i}: \forall \, \Psi_{x_i}\in \mathds{S}_i \right\}$, the following bound holds for the global memory channel $\cM$
\begin{align}\label{eq: boud composed}
F_{\rm CJ}(\mathcal{M},{\rm id}) &\geq (1+\delta)T_{\mathds{S}} - \delta  \\
\text{with}&\qquad   \delta := \max \delta_i
\end{align}
\end{witness}
\noindent 
The derivation can be found in Appendix~\ref{app: prod witness}. It is worth mentioning that when the subsystem bounds can be saturated (e.g.~$\mathds{S}_i$ are $2$-designs), the witness in Eq.~\eqref{eq: boud composed} is tight (there is a channel $\cM$ saturating the inequality), but is not of the form $F_{\rm avg}(\mathcal{M},{\rm id}) \geq T_{\mathds{S}}$. This highlights the fact that the product set $\mathds{S}$ of $2$-designs for the subsystems is not a $2$-design for the global system.\\

As a second construction, we consider a memory for a single $d$-dimensional system,  e.g. the orbital angular momentum of a photon~\cite{ding2015quantum}. As a 2-design is composed of $O(d^2)$ states, when the dimension $d$ is high, preparing and measuring such high number of probe states is very challenging in practice. Thus, it is desirable to construct witnesses which are only based on $O(d)$ states -- the minimal scaling. We now give such a witness using $2d$ probe states taken from two MUBs, namely the computational basis and the Fourier basis, and only performing the same two MUB measurements. It is an application of the methods developed in Ref.~\cite{bavaresco2018measurements}, where high-dimensional entanglement of a bipartite state is certified by bounding its fidelity to a pure target state. Using their fidelity bound and taking these states to be $\varrho^\cM_{AB}$ and $\Phi^+$ provides the following non-linear witness for the memory's fidelity.

\begin{witness}[Two MUB witness~\cite{bavaresco2018measurements}] Consider a measurement in two MUBs $\{B_{a|0}=\Psi_{a,0},B_{a|1}= \Psi_{a,1}\}$ and a source preparing the states $\{\Psi_{a,x}\}$, leading to the observed probabilities
\begin{equation}
    {\rm C}_{a,b}^{x,y} :=  \tr (\cM[\Psi_{a,x}]\,  B_{b|y}).
\end{equation}
The fidelity of the memory channel $\cM$ (in dimension $d$) is bounded by

\begin{align}
    F_{\rm CJ} (\cM, {\rm id})  
    & \geq \frac{1}{d} \sum_b {\rm C}_{b,b}^{0,0} -   \frac{1}{d^2} \sum_{a\neq b} {\rm C}_{a,b}^{1,1} \nonumber \\ &  - \frac{1}{d}\!\!\!\sum_{\substack{a\neq a', a \neq b, \\b \neq b', a'\neq b'}}\!\!\! \gamma_{a,a',b,b'} \sqrt{{\rm C}_{a,b}^{1,1}{\rm C}_{a',b'}^{1,1}}, \\
    \text{with}\quad 
     \gamma_{a,a',b,b'} &= \begin{cases} 1/d & (a-a'-b+b') \text{ mod } d = 0\\
0 &\text{otherwise}\end{cases} \nonumber.
\end{align}
\end{witness}

\noindent Note that other constructions of high-dimensional entanglement witnesses based on fidelity bounds, have been developed recently \cite{Morelli2023,Li2025} and can also be adapted to the memory certification problem.\\

Finally, we ask what is the minimal number of probe states required to faithfully test a quantum memory. For the latter we demand that for an ideal memory, the witness certifies a unit fidelity. Here we answer this question by constructing a witness of a $d$-dimensional memory based only on $d$ probe states, which generalizes the ${\rm T}_{\mathds{S}_2}$ witness we provided in the previous section for the qubit case. 

\begin{witness}[Minimal $d$-state witness] \label{wit:minimal}
    Let $\mathds{S}=\{\ket{\Psi_1},\dots, \ket{\Psi_d}\}$ be a linearly independent set of states, such that $\ket{\Psi_{k+1}}$ is non-orthogonal to the subspace spanned by the states $\ket{\Psi_\ell}$ with $\ell \leq k$, then
    \begin{align} \label{eq:minwitness}
        {\rm T}_{\mathds{S}}= \frac{1}{d}\sum_{x=1}^d F(\Psi_x,\cM[\Psi_x])=1 \iff \cM = {\rm id}.
    \end{align}
\end{witness}
\noindent The proof is given  in Appendix~\ref{app: proof minimal}. Concerning the minimality of this construction, notice that any witness based on less than $d$ states can not be faithful, since such set of probe states does not span the entire Hilbert space. Hence the above construction is minimal, and can already provide relevant bounds on the fidelity. As illustrated in the qubit case, this minimal witness will however severely underestimate the fidelity of the memory in the non-ideal case ${\rm T}_{\mathds{S}}<1$. In this case, computing the actual lower bound on the fidelity can be done by a semi-definite program, as we discuss below.\\

\subsection{General case} \label{sec:sdp}

Let us now turn to a more general scenario going beyond ready-to-use witnesses. Here we show how lower bounds on the average fidelity of the memory can be obtained from generic measurement statistics, by casting the problem as a semidefinite program. To do so, we first need to describe the experiment in a slightly more general context. 

Consider a source producing general quantum states, represented by a set of mixed states (density matrices) $\{ \rho_x \}$. Here we go beyond the case of pure states, which allows e.g.~to include the effect of noise in the probe states. The final measurements are also described in general terms (going beyond projective measurements), namely as POVMs, i.e.~sets of operators $\{B_{b|y} \}$ such that $B_{b|y}\geq 0$ and $\sum_b B_{b|y}  = \openone$ for every $y,b$. The resulting statistics are given by 
\begin{equation}
    {\rm P}(b|\rho_x,y) = \tr( \cM[\rho_x] B_{b|y}) \,.
\end{equation}

From the \ChJa isomorphism it directly follows that
\begin{equation}
    F_{\rm CJ}(\cM,{\rm id}) = \sum_{x,b,y} c_{xby}  {\rm P}(b|\rho_x,y)
\end{equation}
if there exist coefficients $c_{xby}$ such that $ \sum_{x,b,y} c_{xby} \,  \rho_x^\top\otimes B_{b|y} = \frac{1}{d}\Phi^+$, see Appendix~\ref{app: F to F}. More generally, a lower bound on the \ChJa fidelity of the memory channel can be obtained directly from this observed data  via the following semi-definite program~\cite{skrzypczyk2023semidefinite} (SDP)
\begin{align}
       F_{\rm CJ}(\mathcal{M}, {\rm id}) \geq  \min_{\varrho_{AB}} &\quad \tr[\varrho_{AB}\, \Phi^+]  \\
        \text{subject to} &\quad 
            \varrho_{AB} \succeq 0 \nonumber\\
            & \quad d\, \tr_{B}(\varrho_{AB}) = \openone_A \nonumber\\
            & \quad d\, \tr[(\rho_x^\top\otimes B_{b|y})\varrho_{AB}] = {\rm P}(b|\rho_x,y) \nonumber  \,.
\end{align}
Such SDP problems can typically be solved efficiently on standard computers. Moreover, unlike local nonlinear optimization methods (e.g.~those implemented in MATLAB's {\tt fmincon}), algorithms for SDP optimization are guaranteed to converge to the global minimum under mild regularity conditions that are usually satisfied. For more details, see Appendix~\ref{app:SDPopt}.

\subsection{Testing the dimensionality of the memory} \label{sec:dimensionality}

For high-dimensional memories, it is also relevant to certify its ability to transmit high-dimensional quantum information \cite{engineer2025certifying}. That is, we want to ensure that the memory cannot be simulated via a lower-dimensional channel. In the entanglement picture, the dimensionality of the memory is naturally defined via the Schmidt number (i.e.~the entanglement dimensionality) of the Choi state, which we denote ${\rm SN}(\cM)$. The latter is an integer between $1$ and $d$, formally corresponding to the minimal rank of Kraus operators of the channel \cite{chruscinski2006}.

A lower bound on the Schmidt number of the channel is given in terms of the average channel (or Choi) fidelity as ~\cite{terhal2000schmidt}
\begin{align}\label{eq: dimension}
   {\rm SN}(\cM) \geq \lceil d\, F_{\rm CJ}(\mathcal{M},{\rm id}) \rceil = \lceil (d+1)\, F_{\rm avg}(\mathcal{M},{\rm id}) -1 \rceil. 
\end{align}
Hence any lower bound obtained on the channel fidelity, e.g. via the above families of witnesses, directly provides a lower bound on the channel dimensionality.

Finally, note that certifying a Schmidt number of the channel ${\rm SN}(\cM)>1$ ensures that the channel is not entanglement breaking, and leads to the threshold $F_{\rm avg}(\cM,{\rm id})> \frac{2}{d+1}$ mentioned in the Introduction.

\subsection{Testing the quantum capacity of the memory} \label{sec:capacity}

The primary goal of a quantum memory is to transmit quantum information from the past to the future. In quantum information theory the maximal rate at which quantum information can be transmitted through a quantum channel $\cM$ is given by its {\it quantum channel capacity} ${\rm Q}(\cM)$~\cite{lloyd1997capacity,devetak2005private,smith2010quantum}. Operationally, the quantum capacity can be understood as the maximal number of Bell pairs per channel use that can be established between Alice, who controls the channel input, and Bob who receives the channel output.   ${\rm Q}(\cM)$ is an assymptotic quantity which considers the simultaneous usage of $n\to\infty$ copies of the channel\footnote{Formally, one has ${\rm Q}(\cM):=\lim_{n\to \infty}\frac{1}{n}{\rm Q}^{(1)}(\cM^{\otimes n})$}. Nevertheless, it is lower bounded by the so-called single-letter quantum capacity ${\rm Q}^{(1)}(\cM)$, defined in the Appendix~\ref{app: capacity}, which gives the best rate at which quantum information can be transmitted through $\cM$ for independent usage of channel copies.

\begin{figure}
    \centering
    \includegraphics[width=0.9\columnwidth]{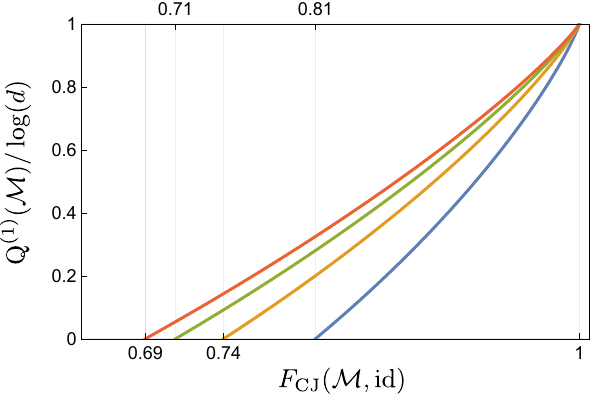}
    \caption{A lower bound on the single-letter quantum channel capacity $Q^{(1)}(\cM)$ normalized by the ideal value $\log(d)$, as a function of the \ChJa fidelity of the channel $F_{\rm CJ}(\cM,{\rm id})$, in dimension $d=2,3,4,5$ (from bottom to top). The numerical values give the threshold fidelity below which no positive quantum capacity is guaranteed.}
    \label{fig: chcap}
\end{figure}

As shown in the Appendix~\ref{app: capacity} the single-letter quantum capacity of the memory can be lower-bounded by an expression involving its Choi state
\begin{align}\label{eq: Q1}
   {\rm Q}^{(1)}(\cM) &\geq  S(\varrho^{\cM}_{B})- S(\varrho^\cM_{AB}) 
   = D(\varrho^\cM_{AB}||\openone_A\otimes \varrho^{\cM}_{B}),
\end{align}
where $\varrho^{\cM}_{B}=\tr_A \left[\varrho^{\cM}_{AB}\right]$, $S(\rho):= -\tr (\rho \log \rho)$ is the von Neumann entropy, and $D(\cdot||\cdot)$ is the quantum relative entropy~\cite{Umegaki1962,Lindblad1975}. Notice that for an ideal memory, the relative entropy in Eq.~\eqref{eq: Q1} is equal to $\log(d)$, and one expects it to increase  with the memory's fidelity $F_{\rm CJ}=F_{\rm CJ}(\cM,{\rm id})$. In the Appendix~\ref{app: capacity} we derive a quantitative relation between the Choi fidelity of the channel and its quantum capacity 
\begin{align}
{\rm Q}^{(1)}(\cM) &\geq \log (d)+F_{\rm CJ} \log (F_{\rm CJ}) \nonumber \\
&+(1-F_{\rm CJ}) \log \left(\frac{1-F_{\rm CJ}}{d^2-1}\right),
\end{align}
 using the joint concavity of the latter~\cite{lieb1973proof}. In Fig.~\ref{fig: chcap} we plot this lower bound for channels $\cM$ in dimension $d=2$ to 5;  while in the large dimension limit $d\to \infty$  it simplifies to ${\rm Q}^{(1)}(\cM) \geq \log(d)(2 F_{\rm CJ}(\cM,{\rm id})-1)$. Notice that there is a regime where the memory is not entanglement-breaking $F_{\rm CJ}(\cM,{\rm id})> \frac{1}{d}$, but is not guaranteed to have a positive quantum capacity. In fact, it is known that the latter condition is strictly stronger~\cite{horodecki2000binding}.

While, in the last two sections we related the fidelity of the channel $F_{\rm CJ}(\cM,{\rm id})$ to its quantum capacity ${\rm Q}^{(1)}(\cM)$ and its dimensionality ${\rm SN}(\cM)$, it is worth mentioning that for a given channel the capacity can be arbitrarily low even in high dimensions. This is because there are channels whose Choi state has full Schmidt number ${\rm SN}(\cM)=d$, but which transmit very low quantum information. A concrete example is the spontaneous decay channel, for which all input states decay to the vacuum state with arbitrarily large probability.

\section{Semi-device-independent scenario}

So far, we have considered a setup where a complete description of the input states and the final measurements is given. In practice, this requires using trusted and pre-calibrated devices. Yet, it is interesting to consider a scenario with a relaxed level of trust, following the so-called semi-device-independent approach \cite{Brask2026}. In the present context, the most relevant point is arguably to relax trust in the device performing the final measurement. This is because, the quantum memory being a complex system, may introduce (unwanted) couplings between the main degree of freedom and others, for example the polarisation of the photons may get coupled to the spatial and/or temporal modes. In turn, this may affect the final measurement statistics and bias the certification process.

\begin{figure}[t]
    \centering
    \includegraphics[width=0.95\linewidth]{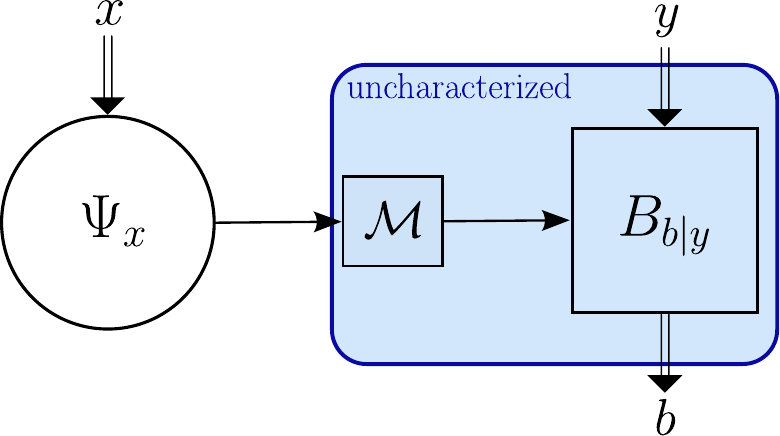}
    \caption{Semi-device-independent scenario for certifying the quantum memory. The preparation device is assumed to be trusted, i.e. upon receiving the classical input $x$, the device produces the state $\Psi_x$ which is then stored in the quantum memory $\mathcal{M}$. The output state of the memory is then measured through a device which is a priori uncharacterized, where $y$ denotes the measurement choice and $b$ its outcome. From the observed statistics, $p(b|\psi_x,y)$, one can lower bound the average channel fidelity of the quantum memory.}
    \label{fig:MDI}
\end{figure}

Here we present a semi-device-independent framework for the certification of a quantum channel. Following the previous discussion we consider a prepare-and-measure scenario, but our methods also apply to an entanglement-based setup. Specifically, we consider that the source produces (trusted) states $\rho_x$ (in $\mathds{C}^d$), while the measurement device is now modeled as a black-box, taking a classical input $y$ and producing a classical binary output $b$. In this setup the quantum memory is modeled as a channel taking as input quantum state in $\mathds{C}^d$ and producing outputs states which are possibly in a much larger Hilbert space $\mathds{C}^{d'}$, with $d' \geq d$. The final measurement is now described by a set of (uncharacterized) operators $B_{b|y}\geq 0$ (such that $\sum_b B_{b|y}  = \openone$), acting on $\mathds{C}^{d'}$. The resulting statistics are given by
\begin{equation} \label{eq: SDI stat}
p(b|\rho_x,y) = \tr( \cM[\rho_x] B_{b|y}) \,.
\end{equation} 
Our goal is now to provide lower bounds on the average channel fidelity based on the observed statistics \eqref{eq: SDI stat}.

In the following, we focus on the qubit case, i.e. $d=2$. Using the set of four Pauli states, $\mathds{S}_4$, we show how to bound the average channel fidelity based on the quantity ${\rm T}_{\mathds{S}_4}$. The two Pauli measurements ($X$ and $Z$) must now be replaced by two arbitrary binary measurements (assumed a priori uncharacterized) denoted $\{B_{b|\updownarrow}\}$ and $\{B_{b|\leftrightarrow}\}$, so that the quantity in Eq. \eqref{eq: T def} now reads
\begin{align}\label{eq: T4 sdi def}
{\rm T}_{\mathds{S}_4} = \frac{1}{4}
\tr \Big[ &B_{0|\updownarrow} \cM[\Psi_1]+ B_{1|\updownarrow} \cM[\Psi_2] \nonumber \\
+&B_{0|\leftrightarrow} \cM[\Psi_3] + B_{1|\leftrightarrow} \cM[\Psi_4] \Big].
\end{align}
with $\Psi_1 = \ket{0}\bra{0}$, $\Psi_2 = \ket{1}\bra{1}$, $\Psi_3 = \ket{+}\bra{+}$ and $\Psi_4 = \ket{-}\bra{-}$. 
In Appendix~\ref{app:MDI}, we demonstrate the following bound
\begin{equation} \label{eq:sdi_bound}
     F_{\rm avg}(\widetilde \cM,{\rm id}) \underset{\rm SDI}{\geq}  
     \frac{1}{3}\left(2(\sqrt 2 +1)(\sqrt{2}\, {\rm T}_{\mathds{S}_4}-1)+1\right)
\end{equation}
where $\widetilde \cM=\Lambda\circ \cM$ is the memory channel composed with {\it an extraction map} $\Lambda$. The bound guarantees that the memory device has the desired quality once the output qubit subspace is properly identified by $\Lambda$\footnote{Note that this extraction map can always realize the trash-and-replace channel, discarding the memory output and preparing the maximally mixed state $\nicefrac{\openone}{d}$. This implies that in the SDI setting the fidelity always fulfills  $F_{\rm avg}(\widetilde \cM,{\rm id})\geq \nicefrac{1}{d}$.}. To derive the bound~\eqref{eq:sdi_bound}, we work in the entanglement picture. Specifically, we consider the Choi state $\varrho^\cM_{AB}$, to which we apply a robust certification technique known as ``self-testing'' \cite{vsupic2020self}. More formally, we consider self-testing methods for the one-sided device-independent (steering-based) scenario~\cite{vsupic2016self}, which we improve by using the extraction map $\Lambda$ introduced in Ref.~\cite{kaniewski2016} for the full DI (CHSH-based) self-testing.

Remarkably, we find that the derived bound is tight for non-entanglement breaking channels. For example, the fully depolarizing channel $\cM : \rho \to (\rho + \sigma_H \rho \sigma_H)/2$, where $\sigma_H =(\sigma_z+\sigma_x)/\sqrt{2}$, has a SDI fidelity $F_{\rm avg}(\widetilde \cM,{\rm id}) = \frac{2}{3}$. As such, it saturates the bound Eq.~\eqref{eq:sdi_bound} as it can reach a witness value ${\rm T}_{\mathds{S}_4} = \frac{1}{4}(2 + \sqrt{2})$ if Bob performs  ${\rm B}_{\leftrightarrow}={\rm B}_{\updownarrow}=\sigma_H$. By convexity, the bound is thus tight for all ${\rm T}_{\mathds{S}_4} \in [\frac{1}{4}(2 + \sqrt{2}),1]$.

Note that these results can also be compared to the fully device-independent setting, where the fidelity of the memory is lower bounded based on the CHSH Bell inequality in the entanglement-based setup~\cite{sekatski2023towards}; see Appendix~\ref{app:FullyDI}. As shown in Fig.~\ref{fig:3}, our SDI analysis performs slightly better.

\begin{figure}
    \centering
    \includegraphics[width=0.95\linewidth]{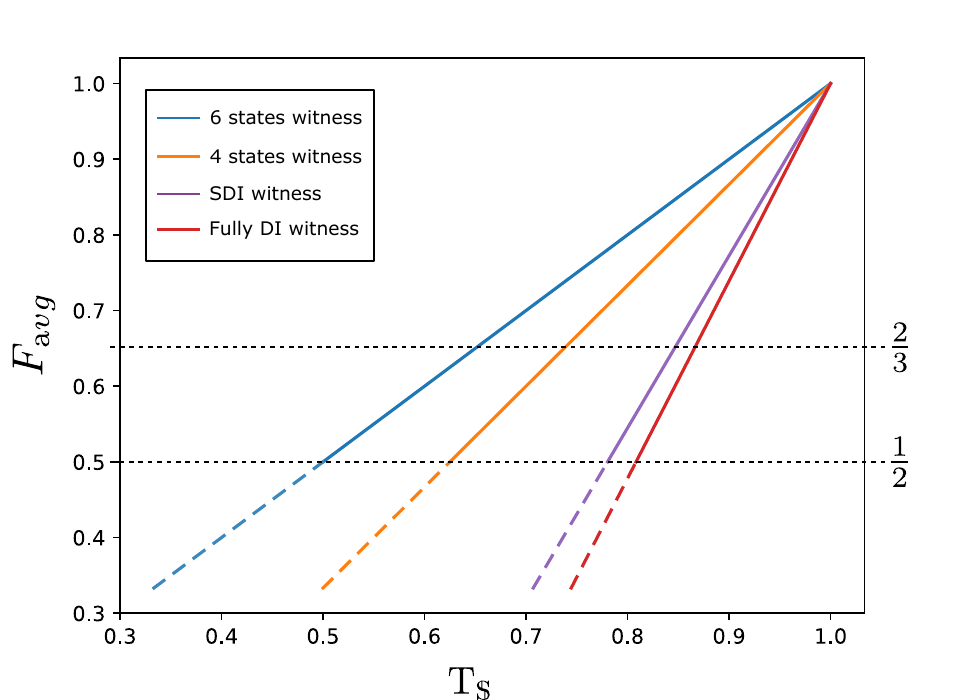}
    \caption{Comparison of the average fidelity one can certify using fully trusted measurements, semi-device-independent certification, and fully device independent (DI) certification. The fidelity in the SDI case, in purple, is certified using the 4 states SDI witness, see Eq.~\eqref{eq:sdi_bound}. In the SDI case, we see that the certifiable fidelity is lower compared to the equivalent 4-state trusted witness. The fidelity in the fully DI case \cite{sekatski2023towards}, in red, is certified if the source characterization is considered perfect, see Appendix~\ref{app:FullyDI}. Note in the SDI and DI case, all channels have minimum SDI fidelity of $1/2$, as the extraction channel can always discard the output state and produce a random two-qubit state.}
    \label{fig:3}
\end{figure}

\section{Experimental application} \label{sec:experiment}

In this section, we present an experimental application of several of the witnesses for the average channel fidelity. The experiment consists of a source of correlated photon pairs and a solid-state $^{171}$Yb$^{3+}$:Y$_2$SiO$_5$ quantum memory. The quantum memory and the source have been described in detail elsewhere \cite{SanchezMejia2025,Rodriguez2026}. Here, we recall some general features of the experimental setup and describe in more detail the specific experiments required for implementing the witnesses proposed in this work.

Quantum storage is achieved with an atomic frequency comb (AFC) memory \cite{Afzelius2009a}, which acts like an effective delay line with a storage time given by the $1/\Delta$, where $\Delta$ is the frequency-domain periodicity of the comb. The memory setup and performance is described in detail in Ref. \cite{SanchezMejia2025}. In this experiment, the storage time was programmed to be $1/\Delta = 10~\mathrm{\mu s}$ and $1/\Delta = 125~\mathrm{\mu s}$. 

The source employs spontaneous parametric down conversion (SPDC) in a non-linear waveguide to produce pairs of photons at 1553 nm and 979 nm, where the latter is bandwidth-matched and actively frequency-stabilized to the absorption line of the quantum memory, see Ref. \cite{Rodriguez2026} for details. The SPDC source is operated in a quasi-continuous mode, producing a coherent stream of pairs of photons in the 1553 nm and 979 modes.  

The qubit storage experiment in this work can conceptually be understood in the following way \cite{Xavier2025}. The detection of a 1553 nm photon (Alice) heralds a single photon in the 979 nm mode (Bob), owing to the strong correlation in the creation time of the two photons within a pair. By weakly pumping the SPDC source, higher-order photon number terms can be suppressed in the 979 nm mode.

The qubit states under consideration here are defined in the time domain, where the computational states $\ket{\pm Z}$ of the $\sigma_Z$ basis correspond to detecting the 1553 nm photon in an early $\ket{E}$ or late $\ket{L}$ time-bin. To herald a superposition state, one can place an unbalanced interferometer in the 1553 nm path. A detection after the interferometer projects the 979 nm photon onto $1/\sqrt{2}(\ket{E} \pm \ket{L})$ or $1/\sqrt{2}(\ket{E} \pm i \ket{L})$, depending on the relative phase of the interferometer, corresponding to projections onto the cardinal states $\ket{\pm X}$ and $\ket{\pm Y}$ of the $\sigma_X$ and $\sigma_Y$ bases. The interferometer is realized with a phase-stabilized fiber-based Michelson interferometer, with a delay of $\tau_d = 19~\mathrm{ns}$ (i.e. the delay between the early $\ket{E}$ and late $\ket{L}$ time bins), see Ref. \cite{Rodriguez2026}. The coherence time of the SPDC pump laser is many orders of magnitude longer, such that the phase coherence of the qubit states are only limited by the interferometer phase stabilization.

The 979-nm photon is analyzed using the same technique. Projections onto the $\sigma_X$ and $\sigma_Y$ basis states are achieved with another interferometer. The 979-nm interferometer was implemented within the quantum memory itself, by storing the qubits with a dual-AFC scheme with a difference in storage time set to $\tau_d$, a common technique that is convenient and requires no additional phase stabilization \cite{Clausen2011}. The corresponding interferometer phase is set by the relative detuning of the two combs, allowing projective measurements onto the $\sigma_X$ and $\sigma_Y$ basis states. Projections onto the computational $\sigma_Z$ basis states correspond to measuring the photon detection time, using a single AFC. Note that both interformeters only have one output port. Hence, the measurements in the $\sigma_X$ and $\sigma_Y$ bases are asymmetric, with a conclusive outcome $E_{\phi} = \eta \ketbra{\phi}$ and an inconclusive  one $\bar E_{\phi} = \openone -\eta \ketbra{\phi}$, where $\ket{\phi}$ is any pure qubit state  (a point on the Bloch sphere) and $\eta$ is the detection efficiency.

\begin{table}[t] 
\centering
\begin{tabular}{l@{\hspace{1.5cm}}c@{\hspace{1.5cm}}c}
\toprule
 Storage time & 10 $\mu$s & 125 $\mu$s \\
\midrule
Efficiency & 16.5\,\% & 3.65\,\% \\
\midrule
6-state estimate of $F_{\rm avg}$ & 98.6\,\% & 98.5\,\% \\
4-state witness of $F_{\rm avg}$  & 98.2\,\% & 98.1\,\% \\
2-state witness of $F_{\rm avg}$  & 78.8\,\% & 78.5\,\% \\
\midrule
SDI witness of $F_{\rm avg}$ & 96.8\,\% & 96.6\,\% \\
\midrule
Quantum capacity ${\rm Q}^{(1)}$ & 81.9\,\% & 80.9\,\% \\
\bottomrule
\end{tabular}
\caption{Experimental results for the average channel fidelity via the different witnesses. Results are given for two different storage times, with corresponding efficiencies. The fidelities are essentially the same for the two storage times, as the memory does not add any significant noise to the output state \cite{Rodriguez2026}. The lower bound on the quantum capacity is obtained from the 6-state estimate. The full measurement statistics are given in Appendix~\ref{app:extra_exp}. 
}
\label{tab: tab1}
\end{table}

To collect the six probabilities appearing in the 2-design witness of Sec.~\ref{sec:perfectqubit}, Alice and Bob align their measurements with the six states $\ket{\phi}\in \{\ket{S_\pm}\}$ from the three Pauli bases $S=X,Y,Z$, always picking the same basis for the measurement. In the experiment, for each basis choice, Alice and Bob record the number of joint conclusive outcomes $N_{++}^S, N_{+-}^S N_{-+}^S$ and $N_{--}^S$. The fidelities for the states $\ket{S_\pm}$ are computed via:

\begin{align} \label{eq:fid_qubits}
 F&\Big(\ketbra{S_\pm}, \cM\big[\ketbra{S_\pm}\big]   \Big) 
    = \begin{cases}
        \frac{N_{\pm,\pm}^S}{N_{\pm,\pm}^S+N_{\pm,\mp}^S} & S\neq Y\\
         \frac{N_{\pm,\mp}^S}{N_{\pm,\pm}^S+N_{\pm,\mp}^S} & S=Y
    \end{cases}.
\end{align}

The measured fidelities for all six probe states $\ket{S_\pm}$ ($S=X,Y,Z$) are given in Appendix~\ref{app:extra_exp}. The average channel fidelity can be evaluated exactly using the 2-design average over all six states, see Eq. \eqref{eq: avg fid sum}. We also apply the two witnesses for obtaining lower bounds on the average channel fidelity based on four and two states, see Eq. \eqref{eq:witnessbounds}, measured in the $X$ and $Z$ bases (cf. the Appendix for details). All the obtained bounds on the fidelities are given in Table \ref{tab: tab1}. In this analysis we assume that the efficiency of the memory is independent of the input state, but can depend on the measurement basis.

The experiment can also be understood in terms of measurements on energy-time entangled photons, using the Franson technique \cite{Xavier2025}, of the type $\ket{\Phi^+}=1/\sqrt{2}(\ket{E_s E_i}+\ket{L_s L_i})$ ($s$ is signal mode at 979 nm and $i$ the idler mode at 1553 nm). In this entanglement-based setting, the average channel fidelity can be evaluated exactly through two-body correlators of the type ${\rm E}_{SS}=\langle \sigma_S \otimes\sigma_S \rangle$, using Eq. \eqref{eq:Corr_fid}, where ${\rm E}_{SS}$ can be computed via:

\begin{align} \label{eq:bp_corr}
   {\rm E}_{SS} 
   & = \frac{N_{+,+}^S+N_{-,-}^S-N_{+,-}^S-N_{-,+}^S}{N_{+,+}^S+N_{-,-}^S+N_{+,-}^S+N_{-,+}^S}.
\end{align}

\noindent The experimental values of ${\rm E}_{SS}$ are given in Appendix~\ref{app:extra_exp}. Note that sometimes experimental groups use the ${\rm E}_{SS}$ correlators and the closely related Choi fidelity $F_{\rm CJ}(\mathcal{M},{\rm id})$, Eq. \eqref{F_CJ_corrs} to calculate the fidelity with respect to the ideal target state $\ket{\Phi^+}$ \cite{Luo2022,zhang2024realization,Zhu2026}, without realizing that the average channel fidelity can be evaluated from exactly the same data.

Finally, the data can also be used to obtain lower bounds on the quantum capacity via Eq. \eqref{eq: Q1}, using the 6-state estimate to bound the Choi fidelity. We also obtain a lower bound on the average channel fidelity for the SDI scenario. In this case, we require a fair-sampling type assumption on the measurement device~\cite{Orsucci2020}, as it is not assumed to be a priori characterized.

\section{Discussion}

We have presented a general and operational framework for quantifying the fidelity of a quantum memory, taking advantage of concepts and tools developed in the theory of quantum channels. The main figure of merit is the average channel fidelity. In practice, this quantity can be efficiently estimated via the concept of $2$-designs. Additionally, we derived a number of witnesses for obtaining lower bounds on the average channel fidelity, which also enable the estimation of other relevant quantities, such as the dimensionality of the channel and its quantum capacity. We discussed these methods for both the prepare-and-measure scenario and the entanglement-based scenario. We also showed that the average channel fidelity can be bounded in a semi-device-independent scenario, where the measurement following the quantum memory is not required to be a priori characterized.

To illustrate their experimental relevance, we have implemented these methods in a solid-state quantum memory experiment. This experiment implements a qubit memory, the average channel fidelity of which is estimated to be  $98.5\%$ via the 2-design protocol which is based on computing the average of the input-output fidelities for the six Pauli eigenstates. Interestingly, the same quantity has already been used in a number of previous quantum memory experiments in order to assess the quality of the memory~\cite{Specht2011single,zhang2011preparation,gundo2012quantum,vernaz2018highly,
Jiang2019,Teller2025,luo2026high}, however, without realizing the connection to the concept of average channel fidelity, and hence rather as an ad hoc procedure. Our work now puts this approach on firm grounds, identifying the precise operational meaning of this quantity, namely the average channel fidelity. 

Similarly, all these tools can be readily applied to an entanglement-based setup, where half of an entangled state is stored in the memory, and the state is finally measured locally. For a qubit memory, the average channel fidelity can be readily estimated from the 3 correlators in the Pauli bases. While this procedure has been followed in previous experiments, our work shows that this approach is well motivated conceptually.

It is also interesting to compare the methods discussed in our work with the protocol of process tomography, see e.g.~\cite{Neeley2008,Specht2011single,Zhou2012}. The latter enables a full reconstruction of the channel, e.g.~in terms of its Choi state, from which the average channel fidelity can also be estimated. However, this protocol is more costly experimentally, requiring the measurement of $O(d^4)$ fidelities. Instead the protocols described here require only $O(d^2)$ fidelities.

Finally, the techniques developed here can be readily used for the certification of other quantum devices, such as optical (fiber) links \cite{neves2025experimentally} and transducers \cite{Bock2024}, the ideal realisation of which also corresponds to the identify channel (as for the quantum memory). Moreover, similar techniques can be developed for characterizing other channels, such as quantum gates and more general unitary operations. Formally, this requires modifying the target channel (or equivalently the target Choi state) and developing corresponding witnesses.

\section{Acknowledgements}

We thank Roope Uola for discussions. The authors acknowledge financial support from the Swiss State Secretariat for Education, Research and Innovation (SERI) under contract numbers UeM019-3 and UeM029-7. This work was also funded with a grant from the Foundation for the University of Geneva (Funige) and the Swiss National Science Foundation (Postdoc.Mobility P500-2-239062).

\bibliographystyle{apsrev4-1}
\bibliography{QM.bib}

\onecolumngrid

\vspace{20pt}
\appendix

\twocolumngrid

\vspace{20pt}
\section{Relation between $F_{\rm avg}$ and $F_{\rm CJ}$ and general fidelity witnesses}\label{app: F to F}

For completeness, we here explain the relation Eq.~\eqref{eq: avg vs CJ} between $F_{\rm avg}(\cM,{\rm id})$ and $F_{\rm CJ}(\cM, {\rm id})$ derived in Refs.~\cite{Horodecki1999,Nielsen2002}. First, we use the \ChJa isomorphism~\cite{Jamiolkowski72} 
\begin{align}\label{eq: Chja}
    \cM[\rho_A] = d\, \tr_B [(\rho_A^\top\otimes \openone_B) \varrho^\cM_{AB}],
\end{align}
where $\varrho^\cM_{AB} =  ({\rm id}\otimes \cM)[\Phi^+]$ is the Choi state of the memory channel, to rewrite the average fidelity of the memory in the entanglement picture
\begin{equation}
    \begin{split}
        F_{\rm avg}(\cM,{\rm id}) &= \int \dd \Psi \tr \left[\Psi\cdot  \cM[\Psi]\right]\\
   &= d \int \dd \Psi \tr \left[(\Psi^\top\!  \otimes \Psi) \varrho^\cM_{AB} \right] \\
   & = d \tr\left[ O\varrho^\cM_{AB} \right],\label{app eq: avg to CJ}
    \end{split}
\end{equation}
with the operator $O:=\int \dd \Psi \left( \Psi^\top\!  \otimes \Psi\right)$.  Interestingly, as we see from the second line, $F_{\rm avg}$ can be understood as the average correlation, observed when measuring the state $\varrho^\cM_{AB}$ in {\it the same random basis} (up to transposition). Entanglement is often presented as synonimous of strong correlations in all measurement basis, and Eq.~\eqref{eq: dimension} relating $F_{\rm avg}$ with the Schmidt number of the state $\varrho^\cM_{AB}$ makes this connection precise.

The operator $O$ is manifestly invariant under the action of $U^{*}\!\otimes U$
\begin{align}
    U^{*}\!\otimes U \,O\, (U^{*}\!\otimes U)^\dag = O,
\end{align}
which can be seen from $U^* \Psi^T (U^*)^\dag = (U \Psi U^\dag)^T$ and the invariance of $\dd \Psi$.
Note that both the identity $\openone$ and the maximally entangled state $\Phi^+$ have this symmetry. In fact, these are essentially the only two such operators~\cite{horodecki1999reduction}, implying the block-diagonal form
$O = x \,\Phi^{+} + \frac{y}{d^2-1}\,(\openone -\Phi^+).$
Since $\tr[O] =1$, we find $y=1-x$, and to determine $O$ it remains to compute 
\begin{align*}
x &=\tr \left[O\, \Phi^+ \right] = \int \dd \Psi \tr[\left( \Psi^\top\!  \otimes \Psi\right) \Phi^+] \\
& =\int \dd U \tr[ \left((U\ketbra{0}U^\dag)^T \!\otimes U\ketbra{0}U^\dag\right) \Phi^+] \\
& =\int \dd U \tr[ U^*\otimes U \ketbra{00} (U^*\otimes U)^\dag  \Phi^+]\\
& = \bra{00}\Phi^+ \ket{00} =\frac{1}{d},
\end{align*}
where $\dd U$ is the Haar measure and we used $U^*\otimes U\ket{\Phi^+}=\ket{\Phi^+}$. We thus find 
\begin{equation}\label{eq: O form}
    O = \frac{1}{d}\left( \Phi^{+} + \frac{1}{d+1}\,(\openone -\Phi^+) \right).
\end{equation}
Plugging this identity into Eq.~\eqref{app eq: avg to CJ} we establish the desired relation 
\begin{align}\label{eq: deriva app a}
   F_{\rm avg}&(\cM,{\rm id}) 
    = d \tr\left[ O \varrho^\cM_{AB} \right] \\
   & = \tr\left[ \left( \Phi^{+} + \frac{1}{d+1}\,(\openone -\Phi^+) \right) \varrho^\cM_{AB} \right] \\
   &= \left(1-\frac{1}{d+1}\right) \tr \left[\Phi^+ \varrho^\cM_{AB}  \right] + \frac{1}{d+1}\\
 &=  \frac{d}{d+1}  F_{\rm CJ}(\cM,{\rm id})+ \frac{1}{d+1}
\end{align}

Let us now consider a general scenario where the measurements $B_{b|y}$ are performed on the states $\rho_x$ stored in the memory, to give the correlations ${\rm P}(b|\rho_x,y)=  \tr (B_{b|y} \cM[\rho_x])$. Using Eq.~\eqref{eq: Chja} we can rewrite any quantity linear in the probabilities as 
\begin{align}
    {\rm T} &= \sum_{x,b,y} c_{xby} {\rm P}(b|\rho_x,y)\\
    &=  d \tr [( \sum_{x,b,y}\, c_{xby}\, \rho_x^T \otimes B_{b|y}) \varrho^\cM_{AB}],
\end{align}
We see that $\rm T$ can be used to faithfully witness the memory's fidelity if the operator 
\begin{equation}
W:= \sum_{x,b,y} c_{xby}\, \rho_x^T \otimes B_{b|y}
\end{equation}
has $\Phi^+$ as the unique maximal (or minimal) eigenstate. Furthermore, we conclude that it is an estimator of the fidelity, i.e. $
{\rm T}=F_{\rm CJ}(\cM,{\rm id}) =\tr (\Phi^+ \varrho^\cM_{AB})$ if 
\begin{equation}
W=  \frac{1}{d} \Phi^{+}.
\end{equation}

\subsection{Trace-non-increasing channels}

Finally, consider a channel $\cM_\checkmark$ that is not trace preserving. In this case we have $\tr \left[\varrho_{AM}^{\cM_\checkmark}\right]= \eta(\cM_\checkmark)$ (see Eq.~\eqref{eq: efficacite} in the main text). Rewriting Eq.\eqref{eq: deriva app a} we thus find 
\begin{align}
   F_{\rm avg}(\cM_\checkmark,{\rm id}) 
   & = \tr\left[ \left( \Phi^{+} + \frac{1}{d+1}\,(\openone -\Phi^+) \right) \varrho^{\cM_\checkmark}_{AB} \right] \\
 &=  \frac{d}{d+1}  F_{\rm CJ}(\cM_\checkmark,{\rm id})+ \frac{\eta(\cM_\checkmark)}{d+1}.
\end{align}
Dividing by the efficiency we find 
\begin{equation}\label{eq app: avg vs CJ}
   (d+1) \bar F_{\rm avg} (\cM,{\rm id})= d \, \bar F_{\rm CJ} (\cM,{\rm id})+1,
\end{equation}
for the conditional fidelities.

\section{Certifying the fidelity of noisy channels}\label{app:SDPopt}

Let's consider a witness in the prepare and measure scenario, involving $n$ states, expressed as 
\begin{equation} \label{eq:appwitness}
    {\rm T}_{\{\Psi_x\}} = \frac{1}{n} \sum_x \Tr(\Psi_x \mathcal{M}(\Psi_x)).
\end{equation}
In the two following subsections, we present ways to analytically and numerically bound the CJ fidelity of the channel $F_{\rm CJ}(\mathcal{M}, {\rm id})$ with respect to a measured witness value ${\rm T}_{\{\Psi_x\}}$. We recall that the CJ fidelity can be directly related to the average fidelity trough Eq.~\eqref{eq: avg vs CJ}. The last two subsections of this appendix discuss respectively optimality in the family of the two states witness for the qubits case, and ways to leverage full statistics information rather than just a single witness value for fidelity certification.

\subsection{Analytical bounds on the fidelity}

Let us first notice that the \ChJa (CJ) fidelity and the witness can be expressed as average values of the Choi state $\varrho^\cM_{AB}:=({\rm id}_A \otimes \mathcal{M}_B)[{\rm \Phi}_d^+]$ of the channel for some well-chosen operators. More precisely, one has
\begin{equation}
    \begin{split}
        F_{\rm CJ}(\mathcal{M}, {\rm id}) = \Tr( {\rm \Phi}_d^+ \varrho^\cM_{AB}), \\
        {\rm T}_{\{\Psi_x\}} = \Tr(W_{\{\Psi_x\}}\varrho^\cM_{AB}),
    \end{split}
\end{equation}
where 
\begin{equation}
    W_{\{\Psi_x\}} = \frac{d}{n} \sum_x \Psi_x^T\otimes \Psi_x.
\end{equation}
This gives us an operational way to bound the $\rm CJ$ fidelity with respect to the witness value. Indeed, the following observation holds:
\begin{equation} \label{eq:linearidbound}
    \begin{split}
        \exists\mu,\nu \in \mathbb{R}, & \ \text{s.t. } {\rm \Phi}_d^+ \succeq \mu W_{\{\Psi_x\}} + \nu \mathds{1} \\
        & \Longrightarrow F_{\rm CJ}(\mathcal{M}, {\rm id}) \geq \mu {\rm T}_{\{\Psi_x\}} + \nu
    \end{split}
\end{equation}
This is due to the fact that $\varrho^\cM_{AB}$ is positive and has trace~$1$. Finding an analytical lower bound to the CJ fidelity thus amounts to proving the existence of such real numbers $\mu,\nu \in \mathbb{R}$. Note that it can even be that those parameters depend on the value of the witness ${\rm T}_{\{\Psi_x\}}$ in order to obtain non-linear lower bounds. Further note that a reasonable property for a good witness is that perfect witness value certifies unit fidelity. As such, any meaningful bound at the optimal value would verify $\mu+\nu=1$. We give below some examples of such bounds.\\

\begin{example}[4-states qubit witness]
     Let us consider the witness $\mathds{S}_4=\big\{\ket{0},\ket{1},\ket{+},\ket{-}\big\}$. We can compute that 
     \begin{equation}
         {\rm \Phi}_2^+ - (2W_{\mathds{S}_4} - \mathds{1}) = {\rm \Psi}_2^- \succeq 0.
     \end{equation}
Using the previous property with $\mu=2$ and $\nu=1$, one has
\begin{equation}
    F_{\rm CJ}(\mathcal{M}, {\rm id}) \geq 2{\rm T}_{\mathds{S}_4} - 1.
\end{equation}
\end{example}

\begin{example}[2-states qubit witness]
    Let us consider the witness $\mathds{S}_2=\big\{\ket{0},\ket{+}\big\}$. One can  check that for all $\mu\in\mathbb{R}$, the operator ${\rm \Phi}_2^+ - (\mu W_{\mathds{S}_2} + (1-\mu)\mathds{1})$ has at least one negative eigenvalue. Thus it is not possible to find an optimal linear lower bound on the CJ fidelity using this method. 
\end{example}

Note that the observation Eq.~\eqref{eq:linearidbound} does not leverage all the information on the Choi state of the channel. Indeed, not only is it trace 1 but taking the partial trace over the first subsystem gives $\mathds{1}/d$ over the second. Therefore, if one denotes $E_{ij}=\mathds{1}\otimes \ket{i}\bra{j}$ the canonical basis on the second system, one has 
\begin{equation} \label{eq:traceconditions}
    \Tr\big(E_{ij}\, \varrho^\cM_{AB}\big) = \frac{\delta_{ij}}{d}.\
\end{equation}
As such one can derive an even tighter linear bound for the fidelity with respect to the witness value:
\begin{equation} \label{eq:linearfullbound}
    \begin{split}
        \exists\mu,\nu \in \mathbb{R},\  & \text{s.t. } {\rm \Phi}_d^+ \succeq \mu W_{\{\Psi_x\}} + \sum_{ij} \nu_{ij}E_{ij} \\
        & \Longrightarrow F_{\rm CJ}(\mathcal{M}, {\rm id}) \geq \mu {\rm T}_{\{\Psi_x\}} + \frac{1}{d}\sum_i \nu_{ii}
    \end{split}
\end{equation}
Finding such bounds can however quite quickly become untractable analytically but can be solved efficiently with a semi-definite programme (SDP). Actually, finding such bound is the dual SDP problem to the minimization of the fidelity with respect to the witness value, as shown in the next subsection.

\subsection{Numerical bounds on the fidelity}

The minimal CJ fidelity, which is achievable by a channel given a witness value ${\rm T}_{\{\Psi_x\}}$, can be computed by the following semi-definite program:
\begin{equation}\label{eq:primalSDP}
    \begin{split}
        \min_{\varrho^\cM_{AB}} & \ F_{\rm CJ}(\mathcal{M}, {\rm id})  \\
        & \text{s.t.} \left\{\begin{split}
            & \varrho^\cM_{AB} \succeq 0 \\
            & \Tr_{\mathcal{H}_2}(\varrho^\cM_{AB}) = \mathds{1}_{\mathcal{H}_1}/d \\
            & \Tr(W_{\{\Psi_x\}}\varrho^\cM_{AB}) = {\rm T}_{\{\Psi_x\}}
        \end{split} \right.
    \end{split}
\end{equation}
Note that the objective function $F_{\rm CJ}(\mathcal{M}, {\rm id}):=\bra{\phi_d^+}\varrho^\cM_{AB}\ket{\phi_d^+}$ is indeed linear in $\varrho^\cM_{AB}$. This numerical method can be more tractable especially for higher dimensional witnesses and can also be used to leverage knowledge about all experimental statistics, i.e.~replacing the last constraint in Eq.~\eqref{eq:primalSDP} with the exact value of each experimental observables, e.g.
\begin{equation}
    \begin{split}
        \min_{\varrho^\cM_{AB}} & \ F_{\rm CJ}(\mathcal{M}, {\rm id})  \\
        & \text{s.t.} \left\{\begin{split}
            & \varrho^\cM_{AB} \succeq 0 \\
            & \Tr_{\mathcal{H}_2}(\varrho^\cM_{AB}) = \mathds{1}_{\mathcal{H}_1}/d \\
            & \forall x, \, d \Tr((\Psi_x^T\otimes \Psi_x)\varrho^\cM_{AB}) = {\rm P} (0| \Psi_x)
        \end{split} \right.
    \end{split}
\end{equation}
see Appendix~\ref{app:biasednoise} for an example. 

Note that the dual SDP problem to Eq.~\eqref{eq:primalSDP} can be written
\begin{equation} 
     \begin{split} 
        \max_{\mu,\nu_{ij}\in \mathbb{R}} & \  \mu {\rm T}_{\{\Psi_x\}} + \frac{1}{d}\sum_i \nu_{ii} \\
        & \text{s.t.} \
            \Phi^+_d - \mu W_{\{\Psi_x\}} - \sum_{ij}\nu_{ij} E_{ij} \succeq 0
    \end{split} \tag{C9dual}
\end{equation}
Hence, any solution to the dual problem does not only give a lower bound on the fidelity but actually a full linear lower bound as it gives coefficients $\mu,\nu_{ij}$ such that Eq.~\eqref{eq:linearfullbound} holds.

\subsection{Product witnesses} \label{app: prod witness}

We here prove the fact that different witnesses for subsystems can be combined for the overall channel as a product witness (Witness~\ref{wit:product}). The proof goes by induction on the number of prime factors of the dimension $d=\prod_{i=1}^n d_i$ of the channel. To do so, we only need to show that, considering two witnesses for dim $d_1$, $d_2$ with 
\begin{equation}
F_{\rm CJ}(\mathcal{M}_i,{\rm id}) \geq (1+\delta_i)T_{\mathds{S}_i} - \delta_i, \end{equation}
the following bound holds  for the product set $\mathds{S}_1 \otimes \mathds{S}_2$ in dim $d:=d_1\times d_2$
\begin{equation}
    F_{\rm CJ}(\mathcal{M},{\rm id}) \geq (1+\delta)T_{\mathds{S}_1 \otimes\mathds{S}_2} - \delta,
\end{equation}
with $\delta = \max \{\delta_1, \delta_2\}$.

As discussed in the previous subsections, the linear lower bounds on the fidelity of each subsystem stem from an operator identity 
\begin{equation} \label{eq:linearopbound1}
    W_{\mathds{S}_i} \preceq \Phi_{d_i}^+ + t_{\mathds{S}_i} \Delta_i,
\end{equation}
where $W_{\mathds{S}_i}$ verifies ${\rm T}_{\mathds{S}_i} = \Tr(W_{\mathds{S}_i}\varrho^{\cM_i}_{AB})$ and some hermitian operator $\Delta$ satisfying $\Delta_i \preceq \openone - \Phi_{d_i}^+$. This indeed leads to the following lower bound:
\begin{equation}
    F_{\rm CJ}(\cM_1, {\rm id}) \geq (1+\delta_i) T_{\mathds{S}} - \delta_i,
\end{equation}
with $\delta_i = t_{\mathds{S}_i}/(1-t_{\mathds{S}_i})$. The global guessing probability $T_{\mathds{S}_1\otimes\mathds{S}_2}$, defined by preparing and measuring each of the two systems stored in the memory independently, verifies that 
\begin{align}
 T_{\mathds{S}_1\otimes\mathds{S}_2}  =    
\tr\left[ W_{\mathds{S}_1}\otimes W_{\mathds{S}_2} \, \varrho^\cM_{AB}\right].
\end{align}
Using Eq.~\eqref{eq:linearopbound1}, we can write
\begin{equation}
\begin{split}
         W_{\mathds{S}_1}\otimes W_{\mathds{S}_2} & \preceq (\Phi_{d_1}^+ + t_{\mathds{S}_1} \Delta_1) \otimes (\Phi_{d_2}^+ + t_{\mathds{S}_2} \Delta_2) \\
        & \preceq (\Phi_{d_1}^+ + t_{\mathds{S}_1}
        (\mathds{1}-\Phi_{d_1}^+))\otimes (\Phi_{d_2}^+ + t_{\mathds{S}_2}(\mathds{1}-\Phi_{d_2}^+))\\ 
        & \preceq {\rm \Phi}_{d}^+ + \max\{t_{\mathds{S}_1}, t_{\mathds{S}_2}\} (\openone -{\rm \Phi}_{d}^+),
    \end{split} 
\end{equation}
where we used $t_{\mathds{S}_1}t_{\mathds{S}_2} \leq \max\{t_{\mathds{S}_1}, t_{\mathds{S}_2}\}$.
Taking the trace over the Choi state $\varrho^\cM_{AB}$ of the memory, we obtain that 
\begin{equation}
    T_{\mathds{S}_1 \otimes\mathds{S}_2} \leq (1-\max\{t_{\mathds{S}_1}, t_{\mathds{S}_2}\}) F_{\rm CJ}(\cM, \openone) + \max\{t_{\mathds{S}_1}, t_{\mathds{S}_2}\}.
\end{equation}
Finally, using the monotonicity of $x\to x/(1-x)$, we indeed have that 
\begin{equation}
    F_{\rm CJ}(\cM, {\rm id}) \geq (1+\delta) T_{\mathds{S}_1 \otimes\mathds{S}_2} - \delta,
\end{equation}
with $\delta = \max\{\delta_1,\delta_2\}$.

A practical example is the one of a memory $\mathcal{M}_N$ storing $N$ identical quantum states. Let $\mathds{S}$ provide a linear witnesses on the average fidelity for a single-system memory $\mathcal{M}_1$, i.e.~
\begin{equation}
    F_{\rm CJ}(\cM_1, {\rm id}) \geq (1+\delta) T_{\mathds{S}} - \delta.
\end{equation}
Then, for the same bound to hold for a witness for the $N$ system memory, consider preparing and measuring each of the systems stored in the memory independently, i.e.
\begin{equation}
    F_{\rm CJ}(\cM_N, {\rm id}) \geq (1+\delta) T_{\mathds{S}^{\otimes N}} - \delta,
\end{equation}
meaning that the original linear lower bound is valid for an arbitrary number of systems $N$. 

In a more general way, this product witness construction gives a general linear witness for arbitrary dim $d$ using the MUBs construction for power-primes~\cite{Wootters1989}. Indeed if $d=\prod_i p_i^{N_i}$ then for each $p_i^{N_i}$ the MUBs provide a perfect witness with $\delta_i=1/p_i^{N_i}$ and thus a $d$-dimensional linear witness can be constructed with $\delta=\max_i 1/p_i^{N_i}$.

\subsection{Proof of the minimal witness}
\label{app: proof minimal}

We now prove that the minimal witnesses (Witness~\ref{wit:minimal}) involving $d$ states for memories of dimension $d$, introduced in Section~\ref{sec:highdim}, are tight in the perfect memory case. Do to so, we consider the family of states $\{\ket{1}, \ket{1+2}, ..., \ket{1+d}\}$, where $\ket{1+k} = (\ket{1}+ \ket{k})/\sqrt{2}$, but the general proof follows from the same logic. We have to verify that if 
\begin{align}
\cM[\ketbra{1}]&=\ketbra{1}\label{eq:111} \tag{r1} \\ 
\cM[\ketbra{1+2}]&=\ketbra{1+2} \tag{r2}\\
&\dots \notag \\
\cM[\ketbra{1+d}]&=\ketbra{1+d}\label{eq:nnn} \tag{rd}
\end{align}
then the channel $\cM$ is the identity. We do that in two steps.

First, combining the identiy (r1) with (rd) one obtains 
\begin{equation}
    \cM[\ketbra{n}{n}+(\ketbra{1}{n}+h.c)] = \ketbra{n}{n}+(\ketbra{1}{n}+h.c).
\end{equation}
Since the channel is trace-preserving, one must have $\cM[\ketbra{n}{n}]= \ketbra{n}+ o$, where $o$ is traceless. Since it's also positive, it follows that $o=0$. Implying, $\cM[\ketbra{1}{n}+h.c] = \ketbra{1}{n}+h.c$. At this point we have
\begin{align}
    \cM[\ketbra{1}] &=\ketbra{1}\\
    \cM[\ketbra{n}] &=\ketbra{n} \\
    \cM[\ketbra{1}{n}+h.c.] &=\ketbra{1}{n}+h.c.
\end{align}
It follows that the channel must act as identity on the qubit subspace $\{\ket{1},\ket{n}\}$: one can probe that $\cM[\sigma_y]$ remains in the qubit subspace (otherwise the channel is not CP), and a CPTP map preserving two Paulis must preserve the third one.

In the second step, we know that the channel acts as identity on the subspaces $\{\ket{1},\ket{2}\}$ and $\{\ket{1},\ket{3}\}$, and show that its must also act as identity on the qutrit space~$\{\ket{1},\ket{2},\ket{3}\}$. 
Let $\Phi^+$ be the maximally entangled two-qutrit state, and define 
\begin{align}
    \rm{R} = \Phi^+ - \frac{1}{3}(\ketbra{22}{33}+\ketbra{33}{22}).
\end{align}
We know that $(\rm{id} \otimes \cM)[\rm{R}] = \rm{R}$, hence it follows that 
\begin{equation}
    (\rm{id} \otimes \cM)[\Phi^+] = \rm{R} + \frac{1}{3}(\ketbra{2}{3}\otimes \cM[\ketbra{2}{3}]+h.c.).
\end{equation}
The operator $\rm{R}$ has a negative eigenvalue $-\frac{\sqrt 2-1}{3}$ assotiated to the eigenstate
\begin{equation}
    \ket{\Psi} = -\frac{1}{\sqrt{2}} \ket{11}+\frac{1}{2}\ket{22} +\frac{1}{2}\ket{33}.
\end{equation}
Since $(\rm{id} \otimes \cM)[\Phi^+]$ is positive, it must be that 
\begin{equation}
    \bra{\Psi}(\ketbra{2}{3}\otimes \cM[\ketbra{2}{3}]+h.c.)\ket{\Psi}\geq \sqrt{2}-1.
\end{equation}
However, this condition is only sufficient to guarantee ${\rm Re} (\bra{2} \cM[\ketbra{2}{3}]\ket{3})\geq 2(\sqrt 2-1)$. Let us rather denote $A=\cM[\ketbra{2}{3}]$, the full positivity condition reads
\begin{equation}
    \rm{R} + \frac{1}{3}(\ketbra{2}{3} \otimes A + \ketbra{3}{2} \otimes A^\dag ) \succeq 0
\end{equation}
where we also know that $A$ is a traceless operator. Note that the operator $\rm{R}$ is supported on $\ket{11},\ket{22},\ket{33}$, so $\ketbra{2}{3}\otimes A$ must also be supported in this subspace (otherwise the full operator can not be positive). This is very restrictive and in fact implies that $A=\alpha \ketbra{2}{3}$ for some complex number $\alpha$. As such, 
\begin{equation}
 (\rm{id} \otimes \cM)[\Phi^+]=   \rm{R} + \frac{1}{3}(\alpha \ketbra{22}{33} +\alpha^*\ketbra{33}{22}).
 \end{equation}
The eigenvalues of this operator are roots of the characteristic polynomial
\begin{align}
    \lambda^3 - 3 \lambda^2 +(1-|\alpha|^2)\lambda +1 +|\alpha|^2 - 2{\rm Re}(\alpha),
\end{align}
and one can verify that the three roots are only positive iff $\alpha =1$, showing that the channel acts as identity on the whole qutrit space.

Iterating this argument we can show that the channel acts trivially on any coherence $\ketbra{i}{j}$ and as the identity on the full $d$ dimensional space.

\subsection{Comparing 2-states witnesses}
When looking at the minimal witnesses in terms of the number of states, one has the degree of freedom of choosing which two states to prepare. The only constrains is that those two states should be linearly independent and have non-zero overlap. When looking at the depolarizing channel of visibility $\lambda$, since its action is invariant under rotation of the Bloch sphere, one can always take the first state to be $\ket{0}$. The only remaining degree of freedom is the angle of the second state with respect to the $z$ axis, and one can write: 
\begin{equation}
    w_2^\alpha = \frac{1}{2}\left( \bra{0} \mathcal{M}(\ketbra{0}) \ket{0} + \bra{u_\alpha} \mathcal{M}(\ketbra{u_\alpha}) \ket{u_\alpha}\right) 
\end{equation}
where $\ket{u_\alpha} := \cos(\alpha) \ket{0} + \sin(\alpha) \ket{1}$.

\begin{figure}
    \centering
    \includegraphics[width=0.95\linewidth]{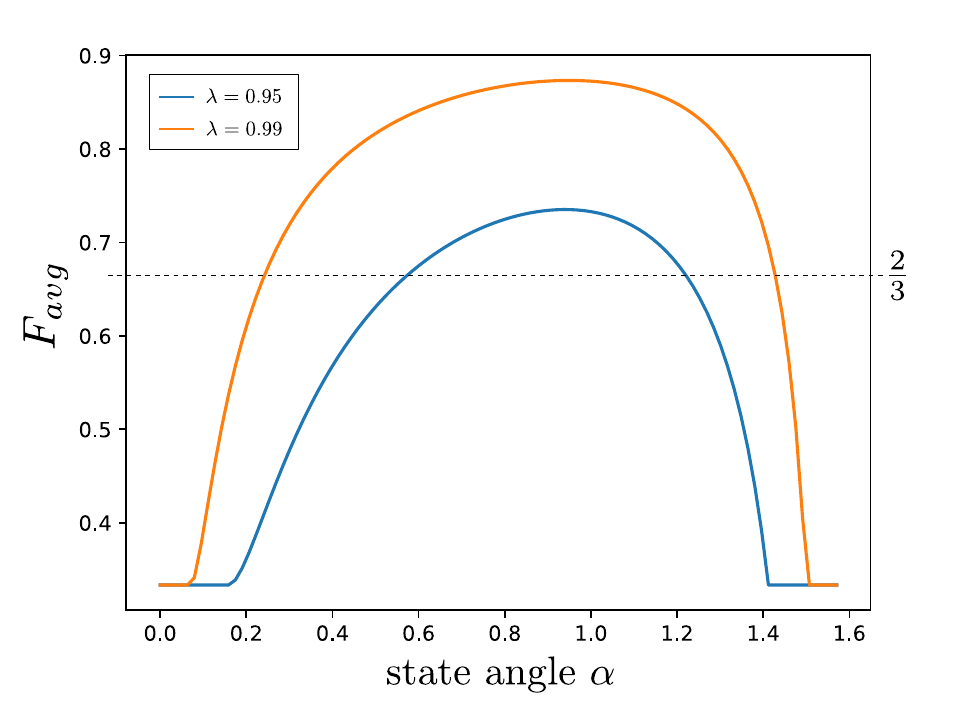}
    \caption{Minimal fidelity to the identity channel that can be certified from a noisy channel that replicates the same witness value $w_2^\alpha$ as a depolarizing channel of visibility $\lambda$. For both visibility values -- $0.95$ and $0.99$ -- the optimal angle is $\alpha^\star \approx 0.95$.}
    \label{fig:fidvsangle}
\end{figure}

The reference value of the witness is always given by $w_2^\alpha(\lambda)^\star = \frac{1}{2}(1+\lambda)$, but the minimal fidelity one can certify for a given $\lambda$ might depend on the value of $\alpha$. In Fig.~\ref{fig:fidvsangle}, we computed the minimal fidelity one can certify depending of the choice of $\alpha$ for different visibility values. 

\subsection{Certifying biased-noise channels} \label{app:biasednoise}

When considering the witness introduced in Eq.~\eqref{eq:appwitness}, a lot of measurements statistics are consumed to compute a single linear quantity. However, in all generality, it grants more information on the noisy channel to look at each measurement statistics independently. This is particularly relevant when the reference noisy channel has a highly biased noise structure. 

\begin{figure}
    \centering
    \includegraphics[width=0.95\linewidth]{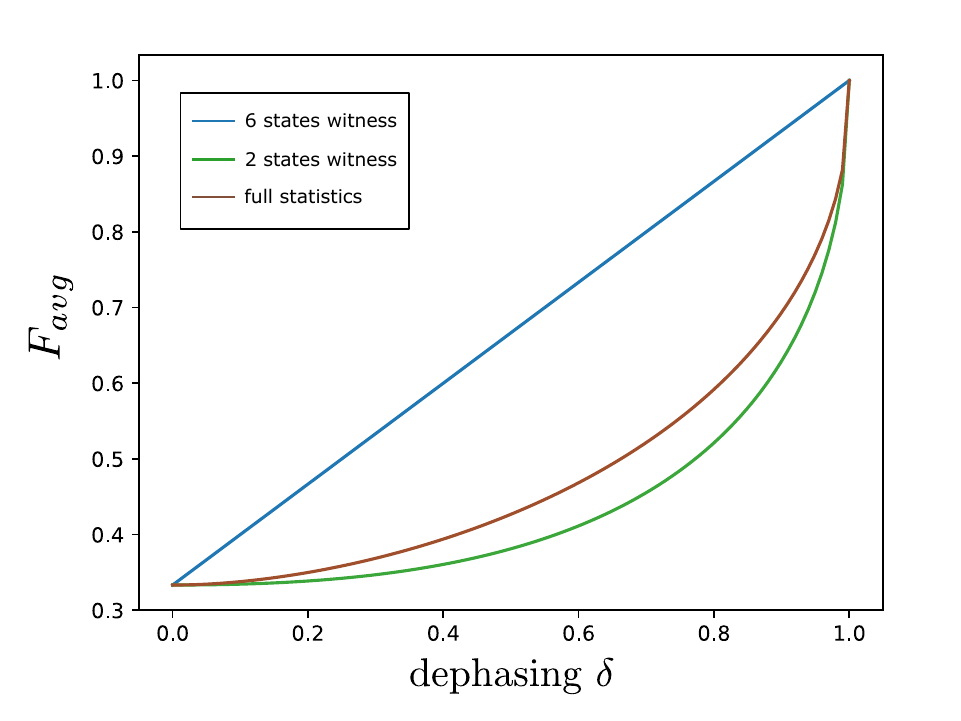}
    \caption{Minimal fidelity to the identity channel that can be certified from a noisy channel that replicates the same witness value that a $z$-dephasing channel $\mathcal{N}_\delta$. The reference fidelity to the identity channel is given by the blue line. The green line is obtained by using the reference value of the 2 states witness with $\ket{0}$ and $\ket{+}$. The brown line is obtained by the full statistics SDP, i.e.~putting both a constraint on both reference values for states $\ket{0}$ and $\ket{+}$, see Eq.~\eqref{eq:fullsdp}.}
    \label{fig:biased-noise}
\end{figure}

As an example, we now consider the reference noisy channel to be a dephasing channel along the $z$ axis of the Bloch sphere, i.e.
\begin{equation}
    \mathcal{N}_\delta [\rho] = \delta \rho + (1-\delta)\sigma_z \rho \sigma_z.
\end{equation}
This noisy channel can be seen as a quantum memory that only showcase phase-flip errors. The value of the observed statistics highly depends on the state that has been prepared. In particular, since the channel leaves the $z$ axis of the Bloch sphere untouched, one has
\begin{equation}
    \bra{0} \mathcal{N}_\delta (\ketbra{0}) \ket{0} = \bra{1} \mathcal{N}_\delta (\ketbra{1}) \ket{1} = 1.
\end{equation}
If one would use the 2-states witness preparing $\ket{0}$ and $\ket{+}$ one would expect a reference value of $(1+\delta)/2$ which is way less binding than imposing that the $z$ axis is preserved while the single state witness value for the $\ket{+}$ state is given by $\bra{+} \mathcal{N}_\delta (\ketbra{+}) \ket{+} = \delta$. The SDP optimization in this case is given by 
\begin{equation} \label{eq:fullsdp}
    \begin{split}
        \min & \ \bra{\phi^+}\varrho^\cM_{AB}\ket{\phi^+}  \\
        & \text{s.t.} \left\{\begin{split}
            & \varrho^\cM_{AB} \succeq 0 \\
            & \Tr_{\mathcal{H}_2}(C_\mathcal{M}) = \mathds{1}_{\mathcal{H}_1}/2 \\
            & 2\bra{00} \varrho^\cM_{AB} \ket{00} = 1\\
            & 2\bra{++}  \varrho^\cM_{AB} \ket{++} \geq \delta
        \end{split} \right.
    \end{split}
\end{equation}
The result of both optimizations (the above one and for the 2 state witness) is given in Fig~\ref{fig:biased-noise}. We also plotted the minimal fidelity one can extract from the 4 state witness and looking at the exact value of each $4$ states statistics. This shows that taking into account both witness values can certify more fidelity while having access to the exact same experimental statistics. 

Note that the same reasoning could be applied to the SDI case by not only looking at the CHSH value but at both the $XX$ and $ZZ$ correlators. In a fully device-independent context, this fine-tuned analysis has shown to be able to certify more state fidelity when considering biased-noise~\cite{Valcarce22}.

\section{Quantum channel capacity and the connection to the \ChJa fidelity.}
\label{app: capacity}

The single-letter quantum capacity of a channel~\cite{lloyd1997capacity,devetak2005private,smith2010quantum} is given by
\begin{equation}\label{eq: cahnnel cap ap}\begin{split}
    {\rm Q}^{(1)}(\cM) &:= \max_\rho I_c(\rho,\cM)\qquad \text{and}\\
    I_c(\rho,\cM) &:= S(\cM[\rho])- S({\rm id} \otimes \cM[\Psi])\\
    &=D\Big(({\rm id} \otimes \cM) [\Psi]\Big|\Big| \openone_A\otimes \cM[\rho]\Big),
\end{split}
\end{equation}
where $\Psi$ is a purification of $\rho$, and $I_c(\rho,\cM)$ is the so-called coherent information~\cite{schumacher1996quantum,lloyd1997capacity}. In words, ${\rm Q}^{(1)}(\cM)$ is the coherent information maximized over all channel inputs. Obviously, a lower-bound on the quantum capacity is obtained if we fix $\rho=\nicefrac{\openone}{d}$ as the completely mixed state, such that $\Psi=\Phi^+$ and $({\rm id} \otimes \cM)[\Psi]= \varrho^\cM_{AB}$ in Eq.~\eqref{eq: cahnnel cap ap} is the Choi state of the memory. We then find
\begin{align}\label{eqapp: Q1}
   {\rm Q}^{(1)}(\cM) &\geq D(\varrho^\cM_{AB}||\openone_A \otimes \varrho_B^{\cM})
\end{align}
with $\varrho_B^{\cM}=\tr_A(\varrho_{AB}^{\cM})$.

To lower bound the quantum capacity of the channel from its fidelity ${\rm Q}^{(1)}(\cM)\geq \mathfrak{q}(F_{\rm CJ}(\cM,{\rm id}))$, we need to solve the following optimization problem
\begin{align}\label{eq: q min}
    \mathfrak{q}(f):= \qquad \min \quad & D(\varrho^\cM_{AB}||\openone_A \otimes \varrho_B^{\cM})\\
    \text{subject to} \quad & F_{\rm CJ}(\cM,{\rm id}) \geq f, \nonumber
\end{align}
or obtain a lower bound on the minimum. To do so, for any channel $\cM$ consider its twirled version 
\begin{align}
    \widetilde \cM := \int \dd\, \mathcal{U} \, (\mathcal{U}^\dag \circ \cM \circ \mathcal{U})
\end{align}
where we apply a Haar-random unitary channel before, and its inverse after, the channel. We now use several properties to rewrite the minimization in Eq.~\eqref{eq: q min}. First, we use the joint convexity of the quantum relative entropy~\cite{lieb1973proof}
\begin{align}
   \sum p_i D(\rho_i ||\sigma_i)\geq  D(\sum_i p_i \rho_i || \sum_i p_i \sigma_i)
\end{align}
and its invariance under unitary transformations $D(\varrho^\cM_{AB}||\openone_A \otimes \varrho_B^{\cM})
     =    D(\varrho^{\mathcal{U}^\dag \circ \cM \circ \mathcal{U}}_{AB}||\openone_A \otimes \varrho_B^{\mathcal{U}^\dag \circ \cM \circ \mathcal{U}})$,
to obtain 
\begin{align}
     &D(\varrho^\cM_{AB}||\openone_A \otimes \varrho_B^{\cM}) \\
     &=   \int \dd\, \mathcal{U}\,   D(\varrho^{\mathcal{U}^\dag \circ \cM \circ \mathcal{U}}_{AB}||\openone_A \otimes \varrho_B^{\mathcal{U}^\dag \circ \cM \circ \mathcal{U}})\\
     &\geq D(\varrho^{\widetilde \cM}_{AB}||\openone_A \otimes \varrho_B^{\widetilde \cM}).
\end{align}
Next, we note that the \ChJa fidelity is invariant under unitary conjugation
\begin{align}
    &F_{\rm CJ}(\mathcal{U}^\dag \circ \cM \circ \mathcal{U},{\rm id}) = \tr (({\rm id}\otimes \mathcal{U}^\dag \circ \cM \circ \mathcal{U})[\Phi^+] \Phi^+)\nonumber \\
    &=\tr (({\rm id}\otimes \cM \circ \mathcal{U})[\Phi^+] ({\rm id}\otimes  \mathcal{U})[\Phi^+]) = F_{\rm CJ}(\cM ,{\rm id}). 
\end{align}
In addition, by linearity we get
\begin{align}
    &F_{\rm CJ}(\widetilde \cM ,{\rm id}) = F_{\rm CJ}\left(\int \dd\,\mathcal{U} \, (\mathcal{U}^\dag \circ \cM \circ \mathcal{U}),{\rm id}\right)\nonumber \\
    &= \int \dd\,\mathcal{U} \,  F_{\rm CJ}\left(\mathcal{U}^\dag \circ \cM \circ \mathcal{U},{\rm id}\right) = F_{\rm CJ}( \cM ,{\rm id}).
\end{align}
We can thus rewrite the minimization in Eq.~\eqref{eq: q min} as
\begin{align}\label{eq: q min 2}
    \mathfrak{q}(f):= \qquad \min \quad & D(\varrho^{\widetilde \cM}_{AB}||\openone_A \otimes \varrho_B^{\widetilde \cM})\\
    \text{subject to} \quad & F_{\rm CJ}({\widetilde \cM},{\rm id}) \geq f. \nonumber
\end{align}
Furthermore, it is intuitive to see that for any $\cM$ the twirled channel $\widetilde \cM $ is always the depolarizing channel
\begin{align}
    \widetilde \cM[\rho] = p \, \rho + (1-p) (\tr \rho) \nicefrac{\openone}{d} 
\end{align}
for some parameter $p\in [0,1]$.
For instance, this can be seen from its symmetry $\widetilde \cM  = \mathcal{U}^\dag \circ \widetilde \cM \circ \mathcal{U}$ for all~$\mathcal{U}$~\cite{nechita2026random}. Hence, it remains to compute $D(\varrho^{\widetilde \cM}_{AB}||\openone_A \otimes \varrho_B^{\widetilde \cM})$ and 
$F_{\rm CJ}({\widetilde \cM},{\rm id})$ as function of the parameter $p$. 

For the Choi state of the depolarizing channel we find
\begin{align}
    \varrho_{AB}^{\widetilde \cM } &= p\,  \Phi^+ +(1-p) (\nicefrac{\openone}{d})\otimes (\nicefrac{\openone}{d})\\
    &= \left(p +\frac{1-p}{d^2}\right)\Phi^+ +\frac{1-p}{d^2}\, \Pi_\perp,
\end{align}
where $\Pi_\perp$ is the projector on the $(d^2-1)$-dimensional subspace orthogonal to $\Phi^+$. This immediately implies 
$F_{\rm CJ}({\widetilde \cM},{\rm id}) = p + \frac{1-p}{d^2}$ and gives the dependence of the parameter $p$ on the fidelity
\begin{align}\label{eq: p of f}
    p=  \frac{d^2 F_{\rm CJ}({\widetilde \cM},{\rm id})-1}{d^2-1}
\end{align}
In turn, to compute the relative entropy we use
\begin{align}
    S(\varrho_{B}^{\widetilde \cM }) &=   S\left(\nicefrac{\openone}{d}\right)=\log(d)\\
    S(\varrho_{AB}^{\widetilde \cM }) & = -\left(p+\frac{1-p}{d^2}\right) \log\left(p+\frac{1-p}{d^2}\right) \\
    &- (d^2-1)\frac{1-p}{d^2} \log (\frac{1-p}{d^2}).
\end{align}
Combining with Eq.~\eqref{eq: p of f} and performing some algebra we find that the coherent information for the depolarizing channels and a fully mixed input state
\begin{align}
I_c(\nicefrac{\openone}{d},\widetilde \cM)&=
   D(\varrho^{\widetilde \cM}_{AB}||\openone_A \otimes \varrho_B^{\widetilde \cM})\\
   &=S(\widetilde \cM[\nicefrac{\openone}{d}])- S(\varrho_{AB}^{\widetilde \cM })
   \\
&= \log (d)+F_{\rm CJ} \log (F_{\rm CJ}) \\
&+(1-F_{\rm CJ}) \log \left(\frac{1-F_{\rm CJ}}{d^2-1}\right).
\end{align}
This concludes the derivation of the bound presented in the main text.

\section{Certifying the fidelity of noisy channels in the SDI scenario} \label{app:MDI}

In this appendix, we prove the lower bound Eq.~\eqref{eq:sdi_bound} on the SDI fidelity using the 4-state witness. In subsection~\ref{app:FullyDI}, we discuss how it relates to the fully DI lower bound. The starting point is to analyze our prepare-and-measure scenario through the corresponding entanglement-based scenario. In the latter, half of a maximally entangled state $\Phi^+$, with $\ket{\Phi^+}=(\ket{00}+\ket{11})/\sqrt{2}$ is stored in the quantum memory. This prepares the bipartite \ChJa state representing the memory channel
\begin{align}\label{eq: CJ def}
  \varrho^\cM_{AB}:= ({\rm id}\otimes\mathcal{M})[\Phi^{+}],
\end{align}
where we denote the two systems as $A$ (auxiliary) and $B$ (memory output). Both systems are then measured, with the probability of any two effects $A_x$ and $B_{b|y}$ given by
${\rm C} (x,b|y) = \tr \left[A_x\otimes B_{b|y} \cdot\varrho^\cM_{AB}\right]$. Then, for the choice $A_x =\Psi_x^\top$ the identity $\tr_A \left[(\Psi_x^\top)_A \otimes \openone_B \cdot \Phi^+_{AB}\right] = \frac{1}{2} (\Psi_x)_B$ establishes the correspondence with the prepare-and-measure scenario ${\rm C} (x,b|y) = \frac{1}{2}\, \tr \left[B_{b|y}\,  \cM[\Psi_x] \right] =\frac{1}{2}\, {\rm P}(b|x,y)$, and allows us to rewrite rewrite Eq.~\eqref{eq: T4 sdi def} it the entanglement-based scenario
\begin{align}\label{eq: ent based 4}
 2\,  {\rm T}_{\mathds{S}_4} -1= \frac{1}{2}&\tr \left[ \left(\sigma_z\otimes {\rm B}_{\updownarrow} + \sigma_x \otimes {\rm B}_{\leftrightarrow} \right) {\tt CJ}[ \cM]\right],
 \end{align}
 where we introduced the observables (Hermitian operators) ${\rm B}_{x}:= B_{0|x}-B_{1|x}$.

Eq.~\eqref{eq: ent based 4} allow us to view our memory certification task as self-testing the post-memory state $\varrho^\cM_{AB}$ from correlations observed in the semi-device-independent (steering) scenario, similar to the setting discusses in e.g.~\cite{vsupic2016self}. Note that since the output of the memory is not trusted (and in particular its dimension could be unknown), to compare the fidelity of the output state to the maximally entangled two qubit state, one first need to apply an{\it extraction map} (i.e.~a quantum channel) $\Lambda_B : L(\mathcal{H}_B) \to L(\mathbb{C}^2)$ on half the output state. This allows for the computation of the fidelity $F((\openone_A \otimes \Lambda_B)[\varrho^\cM_{AB}], \Phi^+)$, being the Choi fidelity in the SDI setting. To compute this, we rely on the fact that only two binary measurements are performed on the uncharacterized system, and use the local extraction map $\Lambda_B$ introduced in Ref~\cite{kaniewski2016} for CHSH-based self-testing. In fact, we prove in the next subsection that 
\begin{equation} \label{eq:lowerboundCJMDI}
         F(({\rm id} \otimes \Lambda_B)[\varrho^\cM_{AB}], \Phi^+_{AB} ) \geq \frac{\sqrt{2}+1}{\sqrt{2}}(2\,  {\rm T}_{\mathds{S}_4} -1)-\frac{1}{\sqrt{2}}.  
\end{equation}
Note that this directly gives the bound~\eqref{eq:sdi_bound} due to Eq.~\eqref{eq: avg vs CJ}.

\subsection{Proof of the lower bound Eq.~\eqref{eq:lowerboundCJMDI}}

To the RHS of Eq.~\eqref{eq: ent based 4}, we associate an operator 
\begin{equation}
    W = \sigma_z\otimes {\rm B}_{\updownarrow} + \sigma_x \otimes {\rm B}_{\leftrightarrow}. 
\end{equation}
Assume that there exists an extraction map $\Lambda_B$ such that this operator is bounded by as 
\begin{equation} \label{eq:Bellopbound}
    ({\rm id} \otimes \Lambda_B^\dagger) [\Phi^+_{AB}] \succeq sW + \mu\, \openone
\end{equation}
where $\Lambda_B^\dagger$ is the dual channel of $\Lambda_B$ and $s,\mu \in \mathbb{R}$. Then the extracted fidelity to the maximally entangled state of any state $\varrho_{AB}$ reaching a score $w=\Tr(\varrho_{AB}W)$ is lower bounded as
\begin{equation} \label{eq:fidbound}
    F(({\rm id} \otimes \Lambda_B)[\varrho_{AB}], \Phi^+_{AB} ) \geq sw+\mu.
\end{equation}
Note that $w=2(2\,  {\rm T}_{\mathds{S}_4} -1)$ due to Eq.~\eqref{eq: ent based 4}. If there exist a family of extractions such that the operator bound Eq.~\eqref{eq:Bellopbound} holds for all measurements choice ${\rm B}_{\updownarrow}$, ${\rm B}_{\leftrightarrow}$, then Eq.~\eqref{eq:fidbound} holds independently of the measurement choice. 

Of course, since the output dimension of the memory is not trusted, parameterizing ${\rm B}_{\updownarrow}$, ${\rm B}_{\leftrightarrow}$ without further restriction is impossible. To circumvent this, we utilize the fact that only two measurements are performed on the system B. In that case, Jordan's lemma applies, and it suffices to show that the previous holds when $\mathcal{H}_B \simeq \mathbb{C}^2$. As such, we consider that system B is a qubit and parametrize the measurements with a single parameter $b \in [0,\pi/2]$ as
\begin{equation}
\begin{split}
    & {\rm B}_{\updownarrow}= \cos(b) \sigma_h +  \sin(b)\sigma_m,\\
    & {\rm B}_{\leftrightarrow}= \cos(b) \sigma_h -  \sin(b)\sigma_m,
\end{split}
\end{equation}
with $\sigma_h =(\sigma_z+\sigma_x)/\sqrt{2}$ and $\sigma_m =(\sigma_z-\sigma_x)/\sqrt{2}$. Note that additional rotations of the qubit space can always be compensated in the extraction $\Lambda_B$. The extraction channel is chosen to be the one introduced in Ref.~\cite{kaniewski2016}:
\begin{equation}
    \Lambda_B(b)[\rho] = \frac{ 1 + g(b) }{2} \, \rho + \frac{ 1 - g(b) }{2} \, \Gamma(b) \, \rho \, \Gamma(b),
\end{equation}
where $g(b) = ( 1 + \sqrt{2}) ( \sin(b) + \cos(b) - 1 )$ and
\begin{align}
\Gamma(b) =
\begin{cases}
\sigma_{h} &\text{for } b \in [0, \pi/4],\\
\sigma_{m} &\text{for } b \in (\pi/4, \pi/2].
\end{cases}
\end{align}
We now proceed to prove that for all $b$:
\begin{equation} \label{eq:opboundapp}
     ({\rm id} \otimes \Lambda_B(b)^\dagger) [\Phi^+_{AB}] \succeq \frac{\sqrt{2}+1}{2\sqrt{2}} W(b) -\frac{1}{\sqrt{2}}{\rm id}, 
\end{equation}
specifying the values of the parameters of Eq.~\eqref{eq:Bellopbound}.

To do so, we introduce the operator
\begin{equation}
    C(b) := K(b) - \frac{\sqrt{2}+1}{2\sqrt{2}} W(b) + \frac{1}{\sqrt{2}},
\end{equation}
with $K(b):=({\rm id} \otimes \Lambda_B(b)^\dagger) [\Phi^+_{AB}]$ and show that all its eigenvalues are positive. Since $C(b)$ is equivalent to $C(\pi/2-b)$ up to a unitary on system B, we only consider wlog $b\in[0,\pi/4]$. We start by writing the operators appearing in the above as
\begin{equation}
    W(b) = \sqrt{2}(\cos(b) \sigma_h \otimes \sigma_h + \sin(b) \sigma_m \otimes \sigma_m),
\end{equation}
and \begin{equation}
    K(b) = \frac{1}{4}(\openone + \sigma_h \otimes \sigma_h+ g(b) (\sigma_m \otimes \sigma_m-\sigma_y \otimes \sigma_y)), 
\end{equation}
since~$\Phi^+_{AB} = \frac{1}{4}(\openone + \sigma_h \otimes \sigma_h+\sigma_m \otimes \sigma_m - \sigma_y \otimes \sigma_y)$. Since only $\sigma_i\otimes \sigma_i$ terms for $i\in\{h,m,y\}$ appear $C(b)$, it is easy to compute its eigenvectors and eigenvalues, respectively:
\begin{subequations}
\begin{align}
    &\ket{\phi_1} = \ket{\phi^+}= (\ket{00}+\ket{11})/\sqrt{2}, \\
    & \ket{\phi_2} = \frac{\ket{\phi^-}+\ket{\psi^+}}{\sqrt{2}}= \frac{\ket{00}+\ket{01}+\ket{01}-\ket{11}}{2}, \\
    & \ket{\phi_3} = \frac{\ket{\phi^-}-\ket{\psi^+}}{\sqrt{2}}= \frac{\ket{00}-\ket{01}-\ket{01}-\ket{11}}{2}, \\
    &\ket{\phi_4} = \ket{\psi^-}= (\ket{01}-\ket{10})/\sqrt{2},
\end{align}
\end{subequations}
and
\begin{subequations}
\begin{align}
    & c_1(b) = 0, \\
    & c_2(b) = (1 + \sqrt{2}) \sin^2(b/2), \\
    & c_3(b) = 1/2\sqrt{2} + (1 + \sqrt{2}) (\cos(b) - \sin(b))/4, \\
    & c_4(b) = 1/2 + (1 + \sqrt{2}) (\cos(b) + \sin(b))/2\sqrt{2}.
\end{align}
\end{subequations}
such that indeed all eigenvalues of $C(b)$ are positive for all $b\in[0,\pi/4]$.

\subsection{Comparison with fully DI case} \label{app:FullyDI}

It is interesting to compare our SDI bound Eq.~\eqref{eq:sdi_bound} with the known fidelity bound in the fully DI setting, based on the CHSH Bell inequality. To make the comparison, we introduce the operators $\sigma_H =(\sigma_z+\sigma_x)/\sqrt{2}$, $\sigma_M =(\sigma_z-\sigma_x)/\sqrt{2}$, and rewrite the RHS of Eq.~\eqref{eq: ent based 4} as
\begin{align}
    \frac{1}{2\sqrt{2}}\tr \left[\left((\sigma_H+\sigma_M)\otimes {\rm B}_{\updownarrow} +(\sigma_H-\sigma_M)\otimes {\rm B}_{\leftrightarrow} \right)\varrho^\cM_{AB}\right].
\end{align}
Replacing the trusted measurement operators with their black-box description $\sigma_H\to {\rm A}_0 = A_{0|0} - A_{1|0}$ with  $\sigma_M \to {\rm A}_1 = A_{0|1} - A_{1|1}$
one precisely recovers the quantum representation of the CHSH Bell test~\cite{Clauser69}. This allows us to interpret the quantity $\mathcal{S}=2\sqrt{2}(2\,  {\rm T}_{\mathds{S}_4}-1)$ as the CHSH score observed for the above choice of observables ${\rm A}_0$ with ${\rm A}_1$. Note that bounding the fidelity in the fully DI case is significantly harder as one needs to also self-test the source (or equivalently characterize the state preparation) without the memory~\cite{sekatski2023towards}. When the source characterization is perfect, the DI fidelity bound reads
\begin{equation}\label{eq:fullyDI}
    F_{\rm avg}(\widetilde \cM,{\rm id}) \underset{\rm DI}{\geq}  
     1 - \frac{4\sqrt{2}}{6 \sqrt{2} - 3 s_\star}(1-{\rm T}_{\mathds{S}_4}),
\end{equation}
where $s_\star = (16 + 14 \sqrt{2})/17 \approx 2.11$. This enables the comparison of the SDI bound~\eqref{eq:sdi_bound} with its DI analogue; see Fig.~\ref{fig:3}.

\section{Detailed experimental results} \label{app:extra_exp}

In Table \ref{tab:av_ch_fidelity_per_basis}, we give for each Pauli basis $S = X, Y, Z$ the correlators ${\rm E}_{SS}$ computed using Eq. \eqref{eq:bp_corr} and the estimated fidelities against each probe state $\ket{S_\pm}$ evaluated using Eq. \eqref{eq:fid_qubits}, as well as the ${\rm T}_{\{\Psi_x\}}$ values that were evaluated to compute the average channel fidelity witnesses given in Table \ref{tab: tab1}.

Note that both interferometers only had a single available output port, so for all phases $(\phi_i, \phi_s)$ corresponding to a measurement in a chosen pair of bases, $(\phi_i + \pi, \phi_s + \pi)$ also had to be measured separately to access the second measurement outcome, meaning that each correlator required 4 separate measurements.

The detection events using two interferometers (the Franson scheme) produce three peaks in the coincidence detection histogram, where only coincidence detections in the middle peak reveal the two-qubit entanglement. The fidelities are computed using the coincidence counts in the middle peak in a time window of 7.65 ns, corresponding to the effective mode duration of the system \cite{Rodriguez2026}.

\begin{table}[h] 
    \centering
   \begin{tabular}{l@{\hspace{1.5cm}}c@{\hspace{1.5cm}}c}
\toprule
Storage time & $10~\mu$s & $125~\mu$s \\
\midrule
${\rm E}_{XX}$ & $0.970$ & $0.965$ \\
${\rm E}_{YY}$ & $-0.969$ & $-0.970$ \\
${\rm E}_{ZZ}$ & $0.976$ & $0.977$ \\
\midrule
$F(X_+)$ & $98.5$\,\% & $98.4$\,\% \\
$F(X_-)$ & $98.5$\,\% & $98.1$\,\% \\
\midrule
$F(Y_+)$ & $98.6$\,\% & $98.7$\,\% \\
$F(Y_-)$ & $98.3$\,\% & $98.3$\,\% \\
\midrule
$F(Z_+)$ & $98.8$\,\% & $98.8$\,\% \\
$F(Z_-)$ & $98.8$\,\% & $98.8$\,\% \\
\midrule
${\rm T}_{\mathds{S}_6}$ & $0.986$ & $0.985$ \\
${\rm T}_{\mathds{S}_4}$ & $0.986$ & $0.985$ \\
${\rm T}_{\mathds{S}_2}$ & $0.986$ & $0.986$ \\
\bottomrule
\end{tabular}
\caption{Experimental results for the $E_{SS}$ correlators, $\ket{S_\pm}$ probe state fidelities $F\Big(\ketbra{S_\pm}, \cM\big[\ketbra{S_\pm}\big] \Big)$, and ${\rm T}_{\{\Psi_x\}}$ witness values. The average standard deviation error on these values is $\pm 0.004$.} \label{tab:av_ch_fidelity_per_basis}
\end{table}

\end{document}